\documentclass[twocolumn]{aastex631}
\usepackage{lipsum}% for placeholder text
\usepackage{amsmath,amstext}
\usepackage{apjfonts} 
\newcommand{\msun}{\ensuremath{M_{\sun}}}

\newcommand{\kms}{\ensuremath{{\rm km~s}^{-1}}}

\newcommand{\ha}{H$\alpha$}
\newcommand{\pb}{Pa$\beta$}
\newcommand{\pg}{Pa$\gamma$}

\newcommand{\hi}{\ion{H}{1}}
\newcommand{\hii}{\ion{H}{2}}
\newcommand{\hei}{\ion{He}{1}}
\newcommand{\ci}{[\ion{C}{1}]}

\newcommand{\nn}{[\ion{N}{1}]}
\newcommand{\nii}{[\ion{N}{2}]}

\newcommand{\oii}{[\ion{O}{2}]}
\newcommand{\oiii}{[\ion{O}{3}]}

\newcommand{\sili}{[\ion{Si}{1}]}

\newcommand{\silx}{[\ion{Si}{10}]}
\newcommand{\pii}{[\ion{P}{2}]}
\newcommand{\si}{[\ion{S}{1}]}
\newcommand{\sii}{[\ion{S}{2}]}
\newcommand{\siii}{[\ion{S}{3}]}

\newcommand{\feii}{[\ion{Fe}{2}]}

\def\simlt{\lower.5ex\hbox{$\; \buildrel < \over \sim \;$}}
\def\simgt{\lower.5ex\hbox{$\; \buildrel > \over \sim \;$}}
\def\herschel{{\it Herschel}}

\def\hst{{\it HST}}
\def\stwofifty{$S_{250~\mu {\rm m}}$}
\def\nhx{$N_{\rm H,\,  X-ray}$}

\def\hei{\ion{He}{1}}
\def\sii{[\ion{S}{2}]}
\def\pii{[\ion{P}{2}]}
\def\feii{[\ion{Fe}{2}]}
\def\siioverfeii {$R$(\sii/\feii)}
\def\piioverfeii {$R$(\pii/\feii)}
\def\heioverfeii {$R$(\hei/\feii)}
\def\piioversii {$R$(\pii/\sii)}

\newcommand{\vl}{\hspace{-1.7em}\vline\hspace{0.em}}

\shorttitle{NIR Spectroscopy of Dense Ejecta Knots in Cas A}
\shortauthors{Koo et al.}

\begin{document}

\title{Near-Infrared Spectroscopy of Dense Ejecta Knots
in the Outer Eastern Area of the Cassiopeia A Supernova Remnant}

\correspondingauthor{Bon-Chul Koo}
\email{koo@snu.ac.kr}

\author[0000-0002-2755-1879]{Bon-Chul Koo}
\affiliation{Department of Physics and Astronomy, Seoul National University,
Seoul 08861, Republic of Korea}
\affiliation{Research Institute of Basic Sciences, Seoul National University, 
Seoul 08826, Korea}
\author[0000-0003-3277-2147]{Yong-Hyun Lee}
\affiliation{Korea Astronomy and Space Science Institute,
Daejeon 305-348, Republic of Korea}
\affiliation{Samsung SDS, Olympic-ro 35-gil 125, Seoul, Republic of Korea}
\author[0000-0003-0894-7824]{Jae-Joon Lee}
\affiliation{Korea Astronomy and Space Science Institute,
Daejeon 305-348, Republic of Korea}
\author[0000-0002-5847-8096]{Sung-Chul Yoon}
\affiliation{Department of Physics and Astronomy, Seoul National University, 
Seoul 08861, Republic of Korea}
\affiliation{Research Institute of Basic Sciences, Seoul National University, 
Seoul 08826, Korea}

\begin{abstract}

The Cassiopeia A supernova remnant has a complex structure,  
manifesting the multidimensional nature of core-collapse supernova explosions. 
To further understand this, we carried out near-infrared multi-object spectroscopy on the 
ejecta knots located in the "northeastern (NE) jet" and 
the ``Fe K plume” regions, which are two distinct features in the outer eastern area of the remnant.
Our study reveals that the knots 
exhibit varying ratios of \sii\ 1.03 \micron, \pii\ 1.189 \micron, and \feii\ 1.257 \micron\ lines 
depending on their locations within the remnant, suggesting regional differences in elemental composition. 
Notably, the knots in the NE jet are mostly `S-rich' with weak or no \pii\ lines, implying that they originated below the explosive Ne burning layer, consistent with the results of previous studies.
We detected no ejecta knots exhibiting only \feii\ lines in the NE jet area
that are expected in the jet-driven SN explosion model.
Instead, we discovered a dozen `Fe-rich' knots in the Fe K plume area. 
We propose that they are dense knots produced by a 
complete Si burning with $\alpha$-rich freezeout in the innermost region of the progenitor and  
ejected with the diffuse X-ray emitting Fe ejecta but decoupled after crossing the reverse shock.
In addition to these metal-rich ejecta knots, several knots emitting only \hei\ 1.083 \micron\ 
lines were detected, and their origin remains unclear.  
We also detected three extended H emission features of circumstellar or interstellar origin in this area 
and discuss its association with the supernova remnant.   

\end{abstract}


\section{Introduction}

The core-collapse supernova (SN) explosion is one of the outstanding problems
in modern astrophysics.
It is generally agreed that, for the majority of core-collapse SNe,  
the explosion is initiated by a core-bounce shock 
which becomes stalled but is revived by neutrino heating. 
Modern state-of-the-art three-dimensional simulations show that
this neutrino-driven SN explosion is basically multi-dimensional, 
facilitated by turbulent convection and hydrodynamic instabilities
\citep[e.g., see][]{janka17,burrows21}.
The non-uniform, inhomogeneous, and 
radially mixed/overturned distribution of metal-rich ejecta 
in SNe and young SN remnants (SNRs) provide 
strong observational evidence for it.

Cassiopeia A (Cas A) is a unique SNR 
where we can see the details of the spatial and velocity distribution of SN ejecta. 
It is young \citep[$\sim 350$ yrs;][]{tho01,fes06} and nearby \citep[3.4 kpc;][]{ree95,ala14}.
It is a remnant of SN Type IIb with a probable progenitor mass of
15--$25~\msun$ (\citealt{young06}; \citealt{krause08}; see also \citealt{koo17} and references therein). The remnant is shell-type and has a basic appearance  
of a small, clumpy bright ring surrounded by a
limb-brightened faint plateau. The bright ring has a radius 
of $1\farcm7$ (1.7 pc) and is mainly composed of 
C and O burning ejecta material heated by reverse shock.
The faint plateau extends to $2\farcm6$ (2.5 pc), and its outer 
boundary defines the current position of the SN blastwave.

In detail, Cas A has a quite complex structure,
manifesting the violent and asymmetric explosion.
In the outer eastern area of the SNR, which is the area of primary interest in this study, 
there are two features distinguishable from the rest of the ejecta material.
One is the ``jet'' structure in the northeastern (NE) 
area and the other is the Fe K-bright ``plume''  
(hereafter the `NE jet' and `Fe K plume', respectively).
The NE jet is a stream of compact optical knots 
confined into a narrow cone well outside the SN blast wave.
It appears to have punched through the main ejecta shell,  
and the inferred ejection velocities is as large as 16,000 \kms, 
which is three times faster than the bulk of the ejecta in the 
main ejecta shell \citep{kamper76,fes96,fes01,fes16}. 
The optical knots show strong O, S, and Ar emission lines 
 \citep{van71,fes96,fes01,fes06,ham08,fes16}. 
The NE jet structure was also observed by {\it Chandra}, 
which showed that the X-ray emitting jet material 
is rich in O-burning and incomplete Si-burning materials (Si, S, Ar, Ca),
but poor in C/Ne-burning products (O, Ne, Mg)
\citep[e.g.,][]{hug00,hwa04,vink04}.
The chemical composition of the jet material  and the jet morphology 
suggest that the NE jet originated from a Si-S-Ar rich layer 
deep inside the progenitor and penetrated through the outer N- and He-rich 
envelope \citep{fes01,fes06,lam06,mil13}.
The other feature is the Fe K plume detected  
by {\it Chandra} and {\em XMM-Newton} in the eastern area
\citep{wil02,hwa03,hwa04,hwa12,tsuchioka22}.
This plume of hot gas bright in Fe K line   
extends beyond the main ejecta shell bright 
in Si K and S K lines, and reaches the forward shock front.
The velocity of the ejecta material in the
outermost Fe K plume is $> 4500$ \kms, 
much higher than that of Si/O-rich ejecta \citep{tsuchioka22}.
The ejecta material in the Fe K plume has no S or Ar, and its Fe/Si abundance 
ratio is up to an order of magnitude higher than
that expected in incomplete Si-burning layer, indicating that they originated from 
complete Si-burning process with $\alpha$-rich freezeout, 
where the burning products are almost exclusively $^{56}$Fe 
(\citealt{thi96,hwa03}, see also \citealt{sato21}). 
\cite{del10} proposed a model for the Fe K plume where 
the Fe ejecta pushed through the Si layer as a ``piston'',  
while \cite{hwa03} and \cite{mil13} suggested that 
$^{56}$Ni bubbles could produce plume-like structures in Fe K emission
but the ionization age of the Fe K plume is inconsistent with a Ni bubble origin.
We will explore the properties 
the NE jet and the Fe K plume in more detail in \S~\ref{sec-dis} 
where we discuss the result of our work in this paper.
Here we only emphasize that 
the distribution of Fe ejecta has not been fully explored: 
The Fe K plume revealed by X-ray observations is 
hot and `diffuse' Fe-rich gas of density $\simlt 10 $ cm$^{-3}$ \citep{hwa03}.
Less hot ($\sim 10^4$~K) and presumably denser Fe-rich ejecta has not been explored. 
The presence or absence of Fe-rich ejecta material in the NE jet is also an important issue related to 
its origin, but it could not have been addressed in the previous optical studies \citep{fes96,fes01,fes16}. 

Recently, \citet{koo18} obtained a long-exposure ($\sim 10$ hr) image of Cas A
by using the UKIRT 4-m telescope with a narrow band filter
centered at \feii\ 1.644~\micron\ emission. 
The band includes \si\ 1.645 \micron\ line, but it is faint and is 
non-negligible only for the unshocked ejecta in the inner area of the SNR 
\citep{koo18,raymond18}, so, 
in this paper, we refer the image as the ``deep \feii\ image''. 
The \feii\ 1.644~\micron\ line is one of the brightest near-infrared (NIR) 
lines in shocked gas, so it traces dense shocked material \citep[e.g.,][]{koo16}. 
The deep \feii\ image revealed numerous knots in the  
outer eastern area beyond the main ejecta shell, 
the proper motion of which indicates that they are mostly SN ejecta material.
The comparison with an {\it HST} WFC3/IR F098M image, which is 
dominated by sulfur lines \citep{ham08,fes16}, showed that 
the majority of the knots in the image correspond to S-rich ejecta knots in optical studies. 
But a considerable fraction (21\%) of the knots does not have optical counterparts, 
and the knots around the Fe K plume in the eastern area  
appear to have high Fe abundance \citep{koo18}.
Hence, the study of these knots could provide new insights on 
the origin of the NE jet and the Fe K plume as well as the explosion dynamics of the Cas A SN.

In this work, we present the results of NIR spectroscopic observations 
of the knots in the deep \feii\ image in the outer eastern area.
In the NIR band (0.95--1.8 \micron), 
in addition to the familiar \ion{He}{1} 1.083 $\mu$m line and 
the forbidden lines of \ion{S}{2}, \ion{S}{3}, and \ion{Fe}{2}, 
there is \pii\ 1.089 $\mu$m ($^{1}D_{2}\rightarrow$$^{3}P_2$) line, 
which is the major NIR line of \ion{P}{2}.
Phosphorus ($^{31}$P) is an uncommon element with cosmic 
abundance relative to H and Fe of  
$2.8\times 10^{-7}$ and $8.1\times 10^{-3}$ by number, respectively 
\citep{asplund09}.
In young SNRs such as Cas A where we can observe SN ejecta, however, 
it could be a major element: 
Phosphorus is produced in hydrostatic neon-burning shells in the pre-SN stage and
also in explosive carbon and neon-burning layers during SN explosion
\citep{arnett96,woosley02}.
Therefore, \pii\ 1.089 $\mu$m line could be strong in some ejecta material 
of young SNRs 
and, together with the other strong lines of $\alpha$ elements (e.g., O, S, Fe), 
it can be used for the study of the SN explosion dynamics \citep{koo13}.
 
 \begin{figure*}
\includegraphics[width=\textwidth]{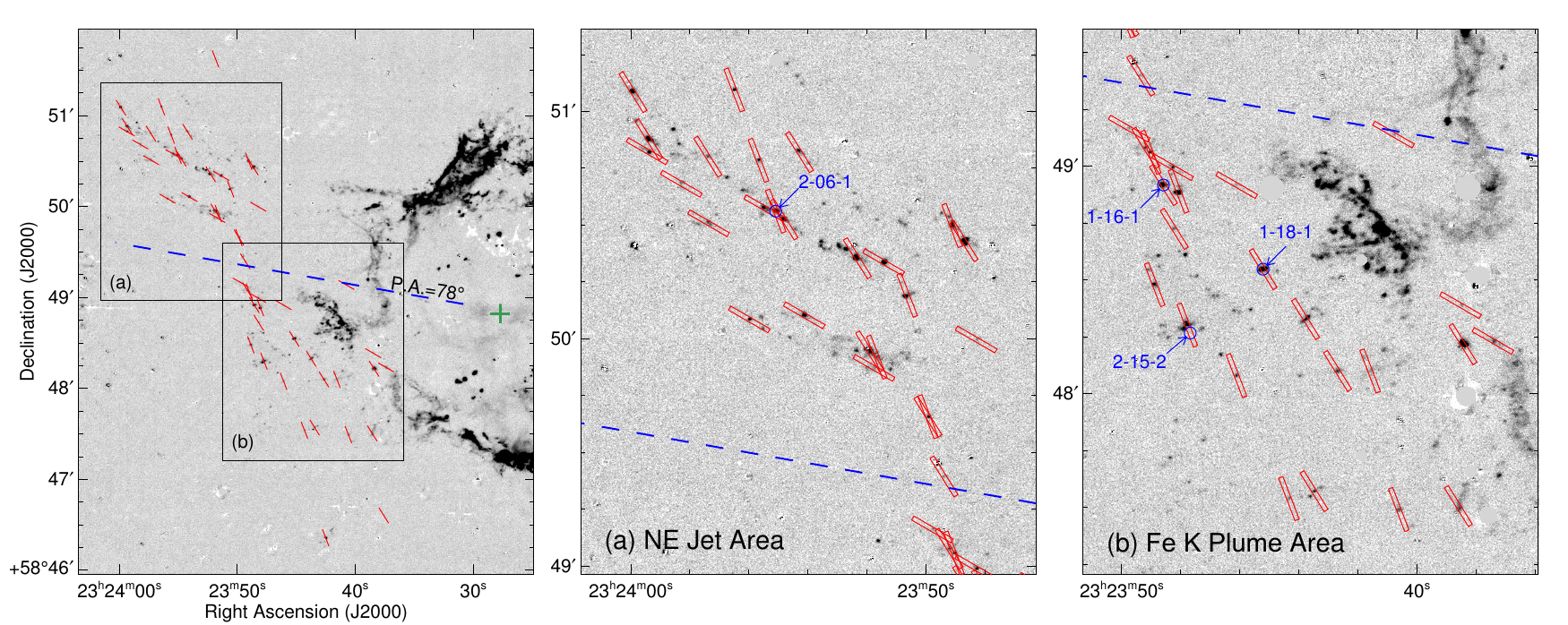}
\caption{
    Finding charts for the slit positions.
    The red bars represent the locations of
    the 10\arcsec-long MOS slits in three masks.
    The background is the deep \feii\ image
    taken in 2013 Sept. \citep{koo18}.
    The green cross marks the SN expansion center,
	($23^{\rm h}23^{\rm m}27\fs77 \pm 0\fs05$,
	$58\degr48\arcmin49\farcs4 \pm 0\farcs4$),
	obtained by \citet{tho01}.
	The blue dashed line represents the position angle of $78\degr$ measured from north to east
	at the explosion center, which is used to divide the knots into two groups:
	(a) NE jet area knots and (b) Fe K plume area knots.
    Note that the slit positions are shifted to the 2013 epoch
    to match the background image (see the text for more detail).
    The 1-D spectra presented in Figure~\ref{fig-1dspec} correspond to the four  
    ejecta knots labeled in blue on the figure. 
} \label{fig-slitpos}

\end{figure*}

The organization of the paper is as follows.
In Section~\ref{sec-obs},
we describe our NIR observations
and explain the data reduction procedures.
In Section~\ref{sec-iden}, we explain 
how we obtain the spectral parameters of the knots for analysis. 
We identify individual knots in two-dimensional spectra 
and extract their one-dimensional spectra. 
We then derive the parameters of individual emission lines 
by performing single Gaussian fits. 
The derived fluxes are extinction corrected by using the 
{\em Herschel} SPIRE 250 $\mu$m data.
In Section~\ref{sec-res},
we analyze the spectral properties of the ejecta knots.
We find that there is significant variation in the ratios of bright lines 
among the knots in different areas. 
We classify the knots into different groups based on the bright line ratios
and compare their spectral properties.
We also search for optical counterparts of the knots and present the result. 
In Section~\ref{sec-dis}, 
we discuss the origins of the knots and 
their connection to the explosion dynamics in Cas A.  We focus 
on the knots in the NE jet region and those in 
the Fe K plume region. 
We also discuss the properties of the extended H emission
features detected in the outer eastern area and its association with the Cassiopeia A SNR.
Finally, in Section~\ref{sec-sum}, we summarize the paper.

\section{Observations and Data Reduction} \label{sec-obs}

We carried out NIR spectroscopic observations
of the SN ejecta knots in the outer eastern area of Cas A in 2017 Sept.
using MMT/Magellan Infrared Spectrograph (MMIRS)
attached on MMT 6.5-m telescope.
MMIRS is equipped with a HAWAII-2RG $2048\times2048$ pixels HgCdTe detector
with a pixel scale of $0\farcs2012$ per pixel, which provides
$6\farcm9\times6\farcm9$ field-of-view for imaging observation \citep{mcl12}.
In order to take $J$ and $H$-band spectra of the ejecta knots,
we utilized the multi-object spectroscopy (MOS) mode
with $J+zJ$ and $H3000+H$ (grism$+$filter) configurations
that provide $J$-band (0.95--1.5\micron)
and $H$-band (1.50--1.80\micron) spectra
with a moderate spectral resolution.

In order to take NIR spectra of the ejecta knots,
we carefully selected the bright knots  in the outer eastern area from 
the deep \feii\ image (Figure \ref{fig-slitpos}).
Since the \feii\ image was taken in 2013,
we estimated the positions of the knots in the 2017 epoch
assuming that they have been expanding freely from the explosion center 
during the age of Cas A.
We prepared three MOS masks, each of which includes 20, 17, 15 MOS slits (Fig. \ref{fig-slitpos}).
Each slit has a width of $0\farcs8$ and a length of $10\arcsec$.
This `moderate' slit width provides the mean spectral resolving power of
$\sim 8$ \AA\ for $J$-band spectra and $\sim 7$ \AA\ for $H$-band spectra,
corresponding to $R\sim1600$ at $1.26~\micron$ and
$R\sim2300$ at $1.64~\micron$, respectively.
At all slit positions in Figure \ref{fig-slitpos}, we obtained $J$-band spectra. 
For the slits in MASK 1, we also took the $H$-band spectra.
In order to subtract sky air-glow emission lines,
we utilized ABBA observing sequence with dithering lengths of
$+2\farcs6$, $-2\farcs0$, $-2\farcs0$, $+2\farcs6$.
The single exposure time per frame was 300 sec,
while the total effective exposure time per mask by the multiple frames
were 40 min or 60 min for the $J$-band spectra,
and 20 min for the $H$-band spectra.
For the absolute flux calibration of the spectra,
we also took the spectra of nearby A0V stars several times
just before or after the target observations.
The seeing during the observation was 0\farcs7--1\farcs2.
The detailed observation log is listed in Table~\ref{tab-log}.

For the data reduction,
we utilized the MMIRS pipeline written in IDL language \citep{chi15}.
The pipeline starts with non-linearity correction of the detector
and cosmic ray removal process,
followed by dark subtraction and flat-fielding correction.
The sky background including bright OH airglow emission lines
was removed by subtracting the dithered frames.
Wavelength calibration and distortion correction were done
by using bright OH airglow emission lines
falling in the $J$ and $H$-band spectra.
The typical $1\sigma$ uncertainty of the wavelength solution is
0.5 \AA\ for $J$-band and 0.3 \AA\ for $H$-band,
corresponding to the velocity uncertainty of $\sim 12~\kms$ at 1.26~\micron\ and
$\sim 5~\kms$ at 1.64~\micron, respectively.
The absolute photometric calibration was performed by applying the correction factor 
as a function of wavelength, derived by comparing the observed spectra of 
standard A0V stars with the Kurucz model spectrum\footnote{\url{http://kurucz.harvard.edu/}}, 
thus it also corrects telluric absorption.
The uncertainty in the absolute photometric calibration has been estimated $\simlt 20$\%, 
while the relative line fluxes in each waveband should be quite robust.

\section{Knot/Line Identification and Extinction Correction} \label{sec-iden}

We clearly detected a total of 67 `knots'
emitting at least one emission line from the 52 MOS slits: 
22 knots in MASK 1, 26 knots in MASK 2, and 19 knots in MASK 3.
No emission line was detected in 4 slits,
while multiple velocity components were detected in 17 slits.
Among the 67 knots, 35 knots are located in the NE jet area
with a position angle (P.A.) of $55\degr$--$78\degr$
from the explosion center,
whereas the rest 32 knots are distributed in the Fe K plume area
with P.A. $>78\degr$ (see Fig. \ref{fig-slitpos}).
Table~\ref{tab-slit} displays the positions of the 52 slits 
where the knots have been detected, and 
Table~\ref{tab-knots} lists the parameters of the identified knots including their radial velocities (see below).

We identified 29 emission lines in total   
in $J$ and $H$-bands, including
forbidden lines from C, Si, P, S, Fe species
and \hei\ 1.083~\micron\ line.
All 67 knots show at least one of the four strongest emission lines;   
\hei\ 1.083~\micron, \sii\ 1.03~\micron\ multiplet, \siii\ 0.983~\micron,
and \feii\ 1.257~\micron\ lines.
About one thirds (27) of them also show strong \pii\ 1.189~\micron\ line,
and they always accompany strong \sii, \siii, and \feii\ lines.
Seven knots show only \hei\ 1.083~\micron\ line
with or without very weak \ci\ 0.985~\micron\ line
above 3--$4\sigma$ rms noise level, while 
five knots show only \siii\ 0.953~\micron\ and/or \sii\ 1.03~\micron\ lines 
(see Table \ref{tab-knots}).
Five out of 22 knots in MASK 1 show \sili\ 1.646~\micron\ line in $H$-band 
together with bright \sii, \siii, \feii, and \pii\ lines.
Previous NIR studies reported 
the detection of \pii\ 1.147 \micron\ and \silx\ 1.430 \micron\ 
lines in the main ejecta shell \citep{ger01,lee17}. In our observations, \pii\ 1.147 \micron\ line is hardly visible from individual spectra, partly because the line is located at the part of spectra with higher noise. We do barely see the line, however, if we stack the spectra of the knots in the NE jet area. 
\silx\ 1.430 \micron\ line mostly falls outside the detector area.

We extracted one-dimensional (1-D) spectra of the 67 knots 
by averaging the spectra of a given region  
and performed single Gaussian fitting for all the detected emission lines
to derive three physical parameters of individual lines:
central velocities, line widths, and line fluxes.
Some emission lines, however, are very close or even blended with each other, so that 
the single Gaussian fitting could not be done without some additional constraints. 
Around 1.08~\micron, for example, there are two strong emission lines from different species,
\si\ 1.082~\micron\ and \hei\ 1.083~\micron,
and their wavelength difference is only 9~\AA.
The spectral resolving power in $J$-band is 8~\AA, so in most cases
we could distinguish whether the line is \si\ or \hei\ from a single Gaussian fitting.
But when the lines are broad (${\rm FWHM}>16$~\AA),
the two lines are blended and  
we had to perform two Gaussian fitting with the central wavelengths and widths of the lines fixed.
For \sii\ 1.03~\micron\ multiplets at  
1.029~\micron, 1.032~\micron, 1.034~\micron\ and 1.037~\micron,
we also fixed their wavelengths and line widths,
as well as their flux ratios based on \citet{tay10}.
Table~\ref{tab-lines} lists the detected lines in the individual knots and their derived parameters.
For the undetected emission lines,
we estimated $1\sigma$ upper limits to their fluxes
using the rms noise level at their expected wavelengths.
All the measured radial velocities have been corrected to
the heliocentric reference frame.

\begin{figure}[!b]
\vspace{1.0truecm}
\begin{center}
\includegraphics[width=0.45\textwidth]{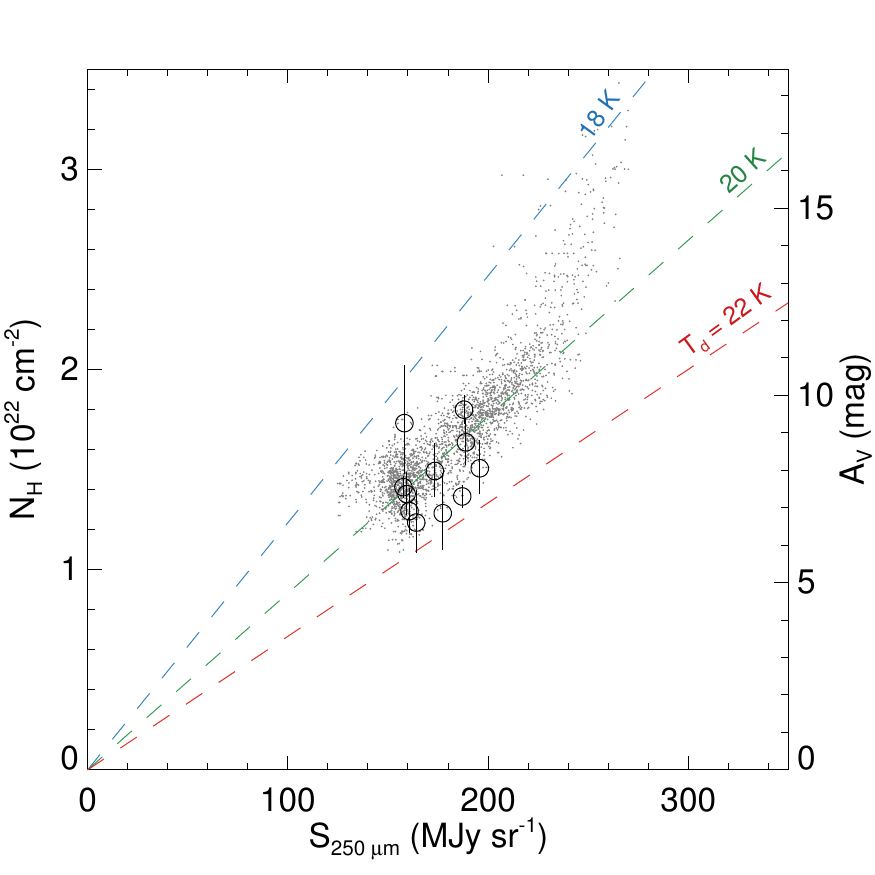}
\caption{
    Hydrogen column density vs. 
    250~\micron\ brightness towards the Cas A SNR. 
    $N_{\rm H}$ is derived from either
    NIR \feii\ line ratios (black empty circles; Section~\ref{sec-iden}) 
    or X-ray analysis \citep[gray dots;][]{hwa12}.
    $S_{250~\micron}$ is obtained from 
    {\it Herschel} SPIRE 250~\micron\ image
    after subtracting synchrotron continuum.
    The dashed lines show the expected $S_{250~\micron}$ of the ISM of 
    $N_{\rm H}$ with dust at $T_{\rm d}=18$, 20, and 22 K.
    The right $y$-axis shows the visual extinction ($A_{\rm V}$) corresponding to
    $N_{\rm H}$ for the general ISM, i.e., $A_{\rm V}/N_{\rm H}=1.87 \times 10^{21}$ mag cm$^{-2}$  
   \citep{dra03}.  
} \label{fig-ext}
\end{center}
\end{figure}

For an analysis of line flux ratios, we need to apply an extinction correction 
to the observed fluxes.
The extinction to Cas A is large 
and varies significantly over the field, e.g., $A_V=5$--15  \citep[][and references therein]{eri09,lee15,koo18}. 
 A column density map of the foreground 
gas/dust across the Cas A SNR had been 
obtained from X-ray spectral analysis \citep{hwa12}, 
but our target knots are mostly outside this column density map. 
For the knots with both \feii\ 1.26~\micron\ and 1.64~\micron\ lines detected, 
we can derive extinction by using the flux ratio of these two lines 
because they share the same upper level \citep[see, e.g.,][]{koo15,lee17}. 
We, however, only obtained $H$-band spectra for the knots in Mask 1, 
and also some knots do not show any detectable \feii\ lines.
So, for the majority of the knots, 
we could not use this technique either.
We instead used the \herschel\ SPIRE 250 $\mu$m data for the extinction correction.
The extinction toward Cas A is mostly due to the ISM in the 
Perseus spiral arm, so that the dust emission at 250 $\mu$m is a 
good measure of the extinction to Cas A \citep[e.g.,][]{delooze17,zhou18}. 
\cite{koo18} showed that there is a good correlation between
the \herschel\ SPIRE 250 $\mu$m surface 
brightness (\stwofifty) and the 
X-ray absorbing column density (\nhx) towards dense circumstellar knots scattered over the 
remnant.
Their plot is reproduced in Figure~\ref{fig-ext},
where we see that the correlation is 
linear at $N_{\rm H,\, X-ray}\simlt 2\times 10^{22}$ cm$^{-2}$ and that it is 
consistent with the 250 $\mu$m emission 
being from 20 K dust with the general ISM dust opacity, i.e., 
$S_{250~\mu {\rm m}}/N_{\rm H,\, X-ray} \approx 1.1~{\rm MJy}/10^{20}~{\rm cm}^{-2}$. 
For higher column densities, \stwofifty\ are considerably smaller than 
those expected from \nhx, which could be either because 
some column densities are due to molecular clouds  
and the temperatures of dust associated with molecular clouds are lower, or 
because some column densities are due to H gas without dust   
\citep{koo18,hartmann97}.
In the figure, we overplotted our target knots where 
both \feii\ 1.26~\micron\ and 1.64~\micron\ lines have been detected 
with high ($>6$) signal-to-noise ratio.
There are 11 such knots and their  \feii\ 1.26 \micron/\feii\ 1.64 \micron\ line ratios range 0.57--0.75 
corresponding to hydrogen column densities of
$N_{\rm H} =1.2$--$1.8\times10^{22}$ cm$^{-2}$ adopting  a theoretical line ratio of 
1.36, which is the value suggested by \citet{nussbaumer88}
and \citet{deb10}, and the dust opacity of the general ISM \citep{dra03}.
We see that the column densities derived from the \feii\ line ratios  
are generally consistent with the 250 $\mu$m brightness
although there is a considerable uncertainty in the former, i.e.,  
the Einstein $A$-coefficients of the two lines are uncertain and 
the \feii\ 1.26 \micron/\feii\ 1.64 \micron\ line ratios in literature range from 0.98 to 1.49 
\citep[see][and references therein]{giannini15, koo15}.
In this work, we have derived the absorbing column densities $N_{\rm H}$ 
for the 67 knots using {\it Herschel} SPIRE 250~\micron\ image 
as described above, and 
applied an extinction correction to the observed line fluxes by using  
the extinction curve of the general ISM \citep{dra03}. 
The derived column densities for individual knots are listed in Table \ref{tab-knots}.
Note that the uncertainty in the extinction correction will not significantly affect 
the results of our analysis using the $J$-band line ratios, 
e.g., an error of $5\times 10^{21}$~cm$^{2}$ in $N_{\rm H}$ would yield 
an error of $\sim 10$\% in the extinction-corrected ratios of \sii\ 1.03 \micron\ and 
\feii\ 1.257 \micron\ line fluxes.

In addition to the 67 knots,
we also detected ``extended H emission features'' in three slits
(2-14, 2-15, 3-18)
in the eastern  outer region with a P.A. of $\sim90\degr$.
They only show hydrogen lines,
%Pa$\gamma$ 1.094~\micron\ and Pa$\beta$ 1.282~\micron,
without any metallic emission lines.
They will be discussed in Section~\ref{sec-dis-ism}, where
we show that they are most likely arising from the 
interstellar/circumstellar medium rather than the SN ejecta.

\section{Result} \label{sec-res}
\subsection{Line Ratios and Their Regional Characteristics}\label{sec-res-flx}

\begin{figure*}
\hspace{0.25in}
\includegraphics[width=0.9\textwidth]{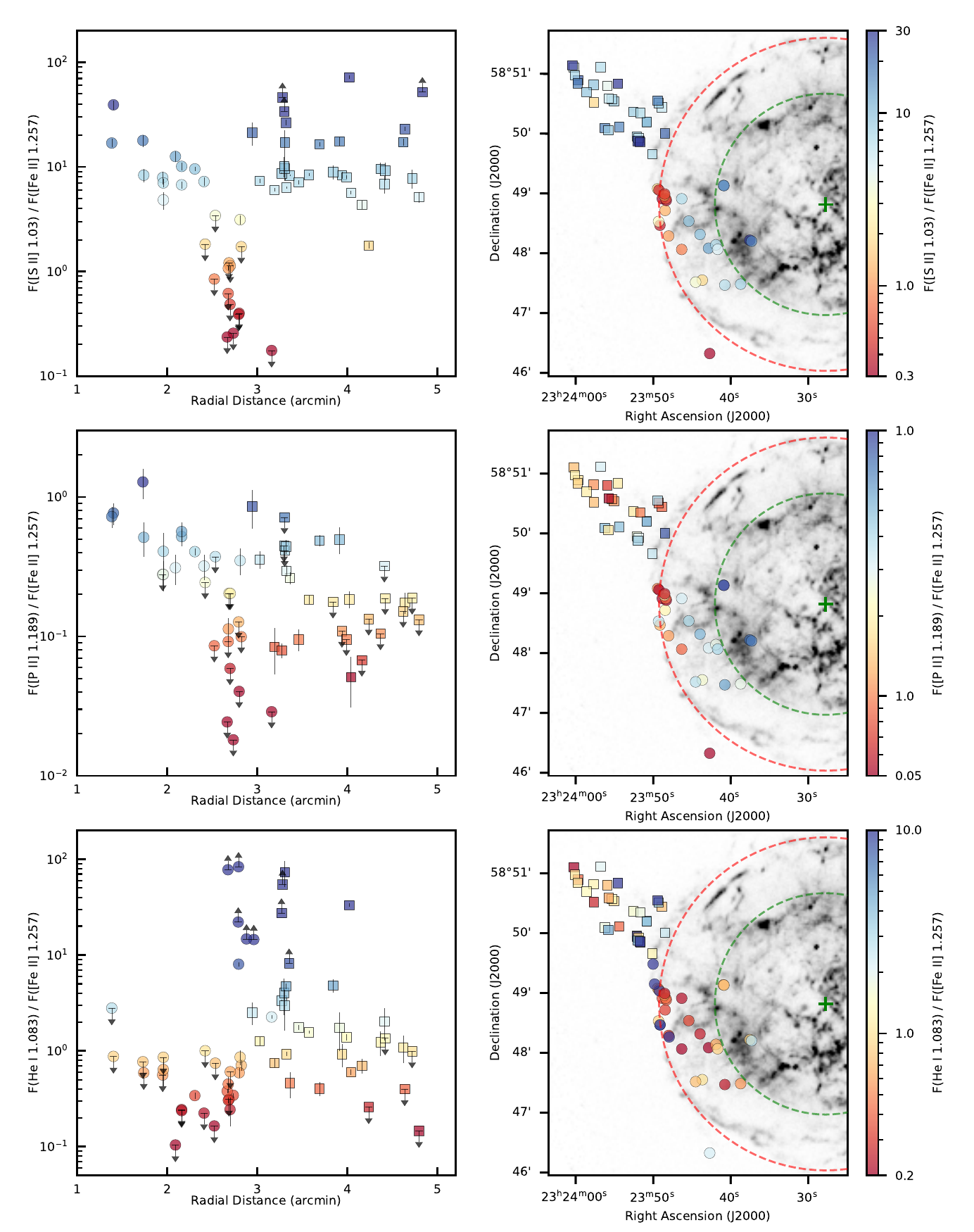}
\caption{
    Identified NIR ejecta knots and their line ratios. 
    The rows represent different line ratios: \siioverfeii\ at the top, \piioverfeii\ in the middle, and \heioverfeii\ 
    at the bottom. The left columns displays the 
    line ratio against the knot’s radial distance from the explosion center, while the right column shows 
    the locations of the knots in the (R.A., Dec.) plane.    
    In all frames, the square and circle symbols indicate the NE jet area and Fe K plume area knots,
   respectively, and their colors represent the flux ratios as in the color bars.  
    The red and green dashed lines in the right frame represent
    the nominal location of the forward and reverse shocks at 2017 epoch, 
    $2\farcm78$ and $1\farcm85$, respectively \citep{got01,del03}.
} \label{fig-flxrat1}
\end{figure*}

Figure~\ref{fig-flxrat1} shows the locations of the observed knots and their 
line flux ratios of three bright emission lines 
(i.e., \sii\ 1.03~\micron, \pii\ 1.189~\micron, and 
\hei\ 1.083~\micron\ lines) to \feii\ 1.257 \micron\ line: hereafter
\siioverfeii, \piioverfeii, and \heioverfeii.
In the left panel, the line flux ratios are plotted as a function of the knot's  
radial distance from the explosion center, while in the right panel, the locations of the 
knots are shown in the (R.A., Dec.) plane.
In all frames, the color of the symbols represents the flux ratios, i.e., red (blue) means low (high) ratios. 
The square and circle symbols indicate the knots in the NE jet and Fe K plume areas, respectively.
The red and green dashed lines in the right panel indicate
the nominal location of the forward and revere shocks, respectively.
We can see that the NE jet knots are located outside of the forward shock,
while most of the knots in the Fe K plume area are located between the forward and reverse 
shocks except one at ${\rm P.A.}=142\degr$ which apparently is located outside the forward shock.

The ratio \siioverfeii\ of the ejecta knots varies over almost two orders of 
magnitude, i.e., 0.2 to 70 (see top row in Figure~\ref{fig-flxrat1}). A noticeable feature is a group of  
knots with low \siioverfeii\ at $r\sim2\farcm7$.
Most of them have a strong \feii\ 1.257~\micron\ line without
any detectable \sii\ 1.03~\micron\ lines, so they are marked with upper limits in the figure.
They are knots located around the forward shock front just  outside the Fe K plume.
Their \siioverfeii\ upper limits are more than one or two orders of magnitude 
lower than the \siioverfeii\ of the knots around the reverse shock at $r=1\farcm4$. 
There also appears to be a systematic variation in the 
\siioverfeii\ of the knots at $r\simlt 2.3$, i.e., 
the knots at larger radial distances have smaller ratios:
The knots at $r=1\farcm4$ have the ratios of a few 10s, while those at 
$r\sim2\farcm3$ have the ratios of $\sim 10$. 
On the other hand, the NE jet knots, except one, have \siioverfeii $> 4$,
which is comparable with those of the knots around the reverse shock.
There is no systematic variation in the \siioverfeii\ of the NE jet knots.

The ratio \piioverfeii\ shows 
a similar pattern to the \siioverfeii, i.e., 
the knots  at $r=2\farcm7$ have a very low \piioverfeii, 
and the knots at smaller radial distances have higher ratios
(see middle row in Figure~\ref{fig-flxrat1}).
Again, most of the knots around $r=2\farcm7$ have
no detectable \pii\ 1.189~\micron\ line.
Most of the NE jet knots also have low \piioverfeii, 
substantially lower than those of the knots around the reverse shock.
Especially, most of the knots at $r \gtrsim 4\arcmin$
do not have any detectable \pii\ 1.188~\micron\ line.
The $2\sigma$ upper limits of these knots are
0.05--0.3, which are a factor of 3--15 smaller than
the ratios of the knots around the reverse shock.

The ratio \heioverfeii\ between the knots differs by a factor of $10^3$ (see bottom row in Figure~\ref{fig-flxrat1}).
For most knots located at $r<2\farcm8$, the \hei\ 1.083 \micron\ line is not detected, 
and the upper limit of the \heioverfeii\ is 0.1--1.0.
However, knots of high ratio abruptly appear at  $r\sim3\farcm0$ and the ratio reaches up to 100. 
Several knots in the NE jet area near $r\sim3\farcm0$ also show exceptionally high
\heioverfeii.
Seven knots were identified that exhibit only \hei\ 1.083~\micron\ line
with/without very weak \ci\ 0.985~\micron\ line. These knots 
will be referred to as ``\hei\ knots'' for the rest of the paper.
These \hei\ knots are detected mainly 
along the forward shock from the southern base of the NE jet to 
the Fe K plume area. It is worth mentioning that the 
radial velocities of the most of the \hei\ knots are very large, so  
they are not circumstellar material (see \S~\ref{sec-dis-se}). 
In the NE jet, the knots near the tip of the jet have low ($\simlt 1.0$) \heioverfeii.

\subsection{Classification of Knots and Their Physical Conditions}
\label{sec-res-cla}
\subsubsection{Knot Classification} \label{sec-res-cla-cla}

In the previous section, we have found that the ejecta knots in 
different areas show distinct NIR spectroscopic properties.  
In this section, we classify the knots by using three line ratios; 
\siioverfeii, 
\piioverfeii, and \heioverfeii. 
We use \sii\ 1.03~\micron\ multiplet line instead of \siii\ 0.953~\micron\ line.
The two lines have a good correlation and 
the latter is usually brighter than \sii\ 1.03~\micron\ multiplet, i.e., 
$F$(\siii\ 0.953)/$F$(\sii\ 1.03) = $1.5\pm0.6$. 
But \siii\ 0.953~\micron\ line is very close to the blue limit of the spectroscope throughput,
so the line becomes hardly detectable if it gets significantly 
blue-shifted (e.g., if its radial velocity is less than $-2000~\kms$).
There are two knots emitting only weak \siii\ 0.953~\micron\ lines, but 
the expected \sii\ 1.03~\micron\ fluxes of those two knots from 
$F$(\siii\ 0.953)/$F$(\sii\ 1.03)=1.5 are smaller than their $3\sigma$ rms noise levels, 
indicating that the non-detection of the \sii\ line is probably due to
the low sensitivity of our observations. 
For these two knots, we use the expected \sii\ 1.03~\micron\ flux 
for their classification for convenience. 

\begin{figure}
\begin{center}
\includegraphics[width=0.45\textwidth]{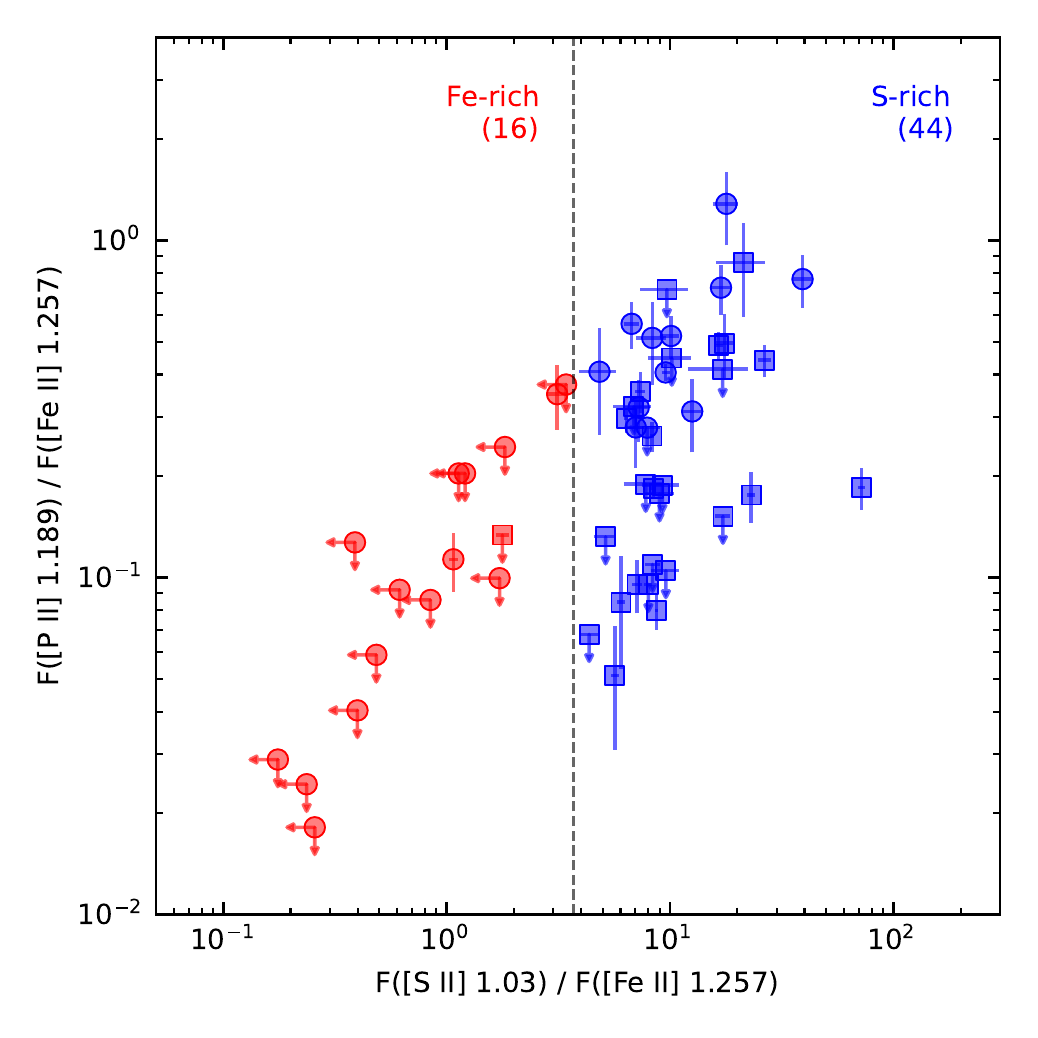}
\caption{Classification of the NIR knots in \siioverfeii\ vs \piioverfeii\ plane. 
   The square and circle symbols indicate the knots in the NE jet and Fe K plume areas,
    respectively. The arrows represent $2\sigma$ upper/lower limits.
    The black dashed line marks the boundary between the 
    S-rich (blue) and Fe-rich (red) knots.
    The number of knots belonging to the two groups are given in parenthesis.
    The number (44) of the S-rich knots includes five knots that could not be not shown in the figure because 
    only \sii\ lines are detected.  
    There are also seven \hei\ knots where these metal lines are not detected (see \S~\ref{sec-res-cla-cla}).    
} \label{fig-class}
\end{center}
\end{figure}

\begin{figure*}
\includegraphics[width=1.0\textwidth]{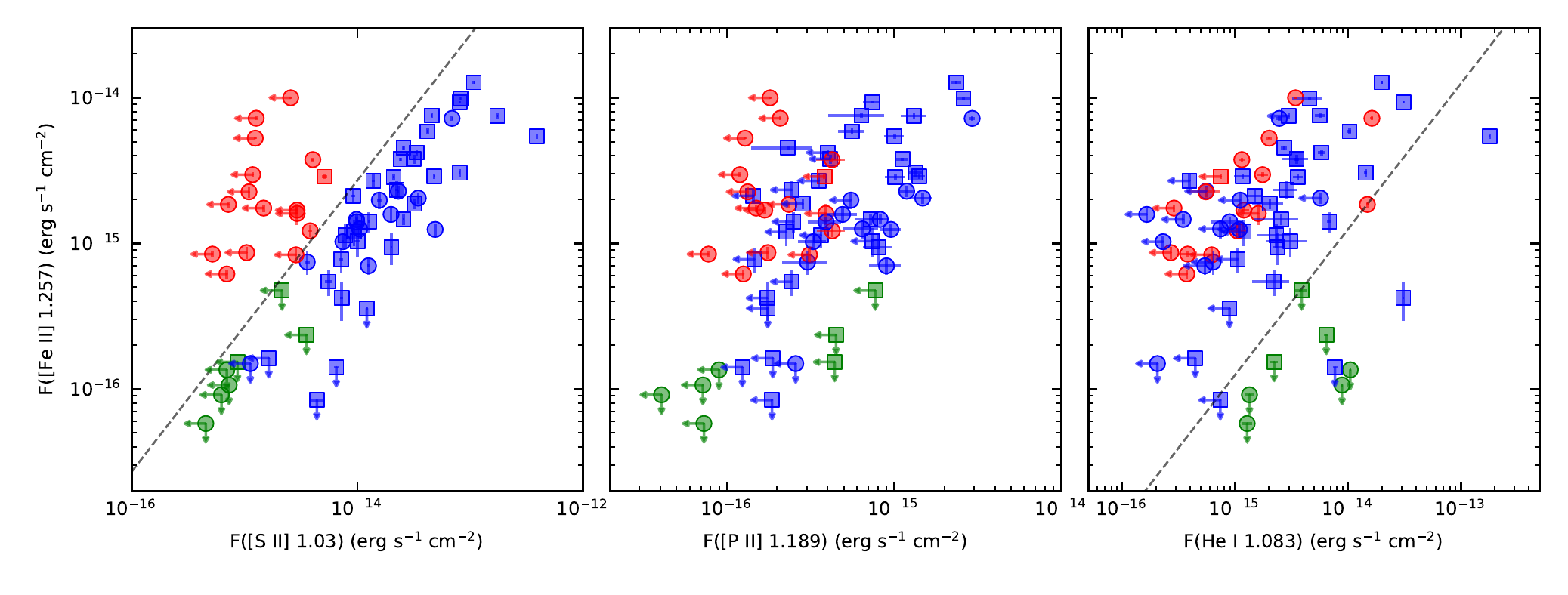}
\caption{
	$F$(\feii) versus $F$(\sii), $F$(\pii), and $F$(\hei). 
	The knot symbols and their red and blue colors are the same as in Figure \ref{fig-class}.
	The green symbols represent \hei\ knots. The arrows represent $2\sigma$ upper/lower limits.
The dashed line in 
	the left frame marks the boundary between S-rich and Fe-rich knots. The dashed line in the right frame 
	is the line at which \heioverfeii=8. The \hei\ knots have 
	 \heioverfeii\ higher than this.
} \label{fig-fxvsfx}
\end{figure*}

Figure~\ref{fig-class} is a diagram of 
\piioverfeii\ versus \siioverfeii.
The NE jet area knots and Fe K plume area knots are marked by squares and circles, respectively.
We first examine \siioverfeii\ of the knots. 
The \siioverfeii\ of the knots ranges from 0.2 to 70.
The Fe K plume area knots are scattered over the entire range, while  
the NE jet knots have relatively high ratios, i.e., \siioverfeii $\simgt 4$.  
Note that most of the Fe K plume area knots with \siioverfeii $\simlt 4$ 
do not have detectable \sii\ emission. Such separation of the knots in 
\siioverfeii\ is similar to the properties of the 
knots in the main ejecta shell. For the  
knots in the main ejecta shell, \cite{lee17} performed principal component analysis 
of their NIR spectral properties and showed that they could be classified 
into two groups; (1) S-rich knots  showing 
strong \sii\ and \pii\ lines and (2) Fe-rich knots showing 
strong \feii\ lines. 
They found that the two groups were well divided  
by the ratio $F$(\sii\ 1.03)/$F$(\feii\ 1.644) of $\sim 5$.
Following \cite{lee17}, we adopt the threshold of \siioverfeii $\sim 3.7$
assuming the intrinsic ratio of 1.36 for 
$F$(\feii\ 1.257 )/$F$(\feii\ 1.644) \citep{nus88,deb10}.
We therefore divide the knots into two groups; 
``S-rich knots'' with \siioverfeii\ $\ge 3.7$
and ``Fe-rich knots'' with \siioverfeii\ $< 3.7$.
In Figure \ref{fig-class}, S-rich and  
Fe-rich knots are represented by blue and red symbols, respectively.
Note that all Fe-rich knots except one are the knots in the Fe K plume area
and that almost of all of them have no detectable \sii\ and \pii\ emission.
The Fe-rich knots are mostly located around 
the outer boundary of the Fe K plume (see Figure \ref{fig-flxrat1}), 
and, as we will discuss in \S~\ref{sec-dis-se}, 
it has an interesting implication on the SN explosion dynamics.  
Another thing to note is that 
all NE jet knots except one are S-rich knots
and that they have \piioverfeii\ ratios generally lower than those of the S-rich knots in the 
Fe K plume area (see also Figure \ref{fig-flxrat1}). 
We will explore the spectral properties of the NE jet knots in detail in \S~\ref{sec-dis-ne}.  

Figure \ref{fig-class} does not include all 67 ejecta knots
because some knots do not exhibit \feii\ 1.257~\micron\ (and \pii\ 1.189~\micron) lines. 
This may appear strange because our target knots have been  
selected from the deep \feii\ image. But the observation was done with a 
$10''$ slit in ABBA mode, so there could be non-\feii-emission knots accidentally located within 
the slit. 
There are 12 knots with no detectable \feii\ 1.257~\micron\ lines.
Among these, four knots show only \sii\ 1.257~\micron\ and/or 
\siii\ 0.953 \micron\ lines, 
one knot shows only \sii\  1.257~\micron\ and \hei\ 1.083~\micron\ lines, and
seven knots show only \hei\ 1.083~\micron\ lines. 
The former five knots belong to S-rich knots  
because \feii\ 1.257~\micron\ line has not been detected. (Their $2\sigma$ lower limits of 
\siioverfeii\ are $\ge 3.7$.)
But \piioverfeii\ is not available for them because both  
\feii\ 1.257~\micron\ and \pii\ 1.189~\micron\ lines are not detected.
Then there are seven  \hei\ 
knots that show only \hei\ 1.083~\micron\ 
line with no \sii\ 1.03~\micron, \feii\ 1.257~\micron, and \pii\ 1.189~\micron\ 
lines. Their \hei\ 1.083~\micron\ fluxes are comparable to those of the metal-rich ejecta knots, 
but their \sii\ 1.03~\micron, \feii\ 1.257~\micron, and \pii\ 1.189~\micron\
fluxes are considerably lower than those of the other ejecta knots (Figure \ref{fig-fxvsfx}). 
We note that these \hei\ knots are mostly an accidental detection and that 
the category of the \hei\ knots is not very rigorously defined. 
Indeed some spectra of S/Fe-rich knots with high \heioverfeii\, 
if degraded, could become indistinguishable from the \hei\ knots. 
However, the number of such knots is less than a few thus should not affect 
our discussion on \hei\ knots.
From Figure \ref{fig-fxvsfx}, we can also see that 
S-rich and Fe-rich knots have a comparable range of \feii\ 1.257~\micron\ line strengths. 
This is not unexpected, 
given that the knots that we observed are the bright ones in the deep \feii\ image.
The two groups of the knots, however,  
are clearly distinguished in their \sii\ line fluxes.
In \pii\ emission, the Fe-rich knots generally have lower 
1.189~\micron\ fluxes than the S-rich knots, but not necessarily. 
A good fraction of the S-rich knots have \pii\ 1.189~\micron\ fluxes as low as the Fe-rich knots.
As we will see in \S~\ref{sec-dis-ne}, they are the knots 
near the tip of the NE jet.

\begin{figure*}
\includegraphics[width=\textwidth]{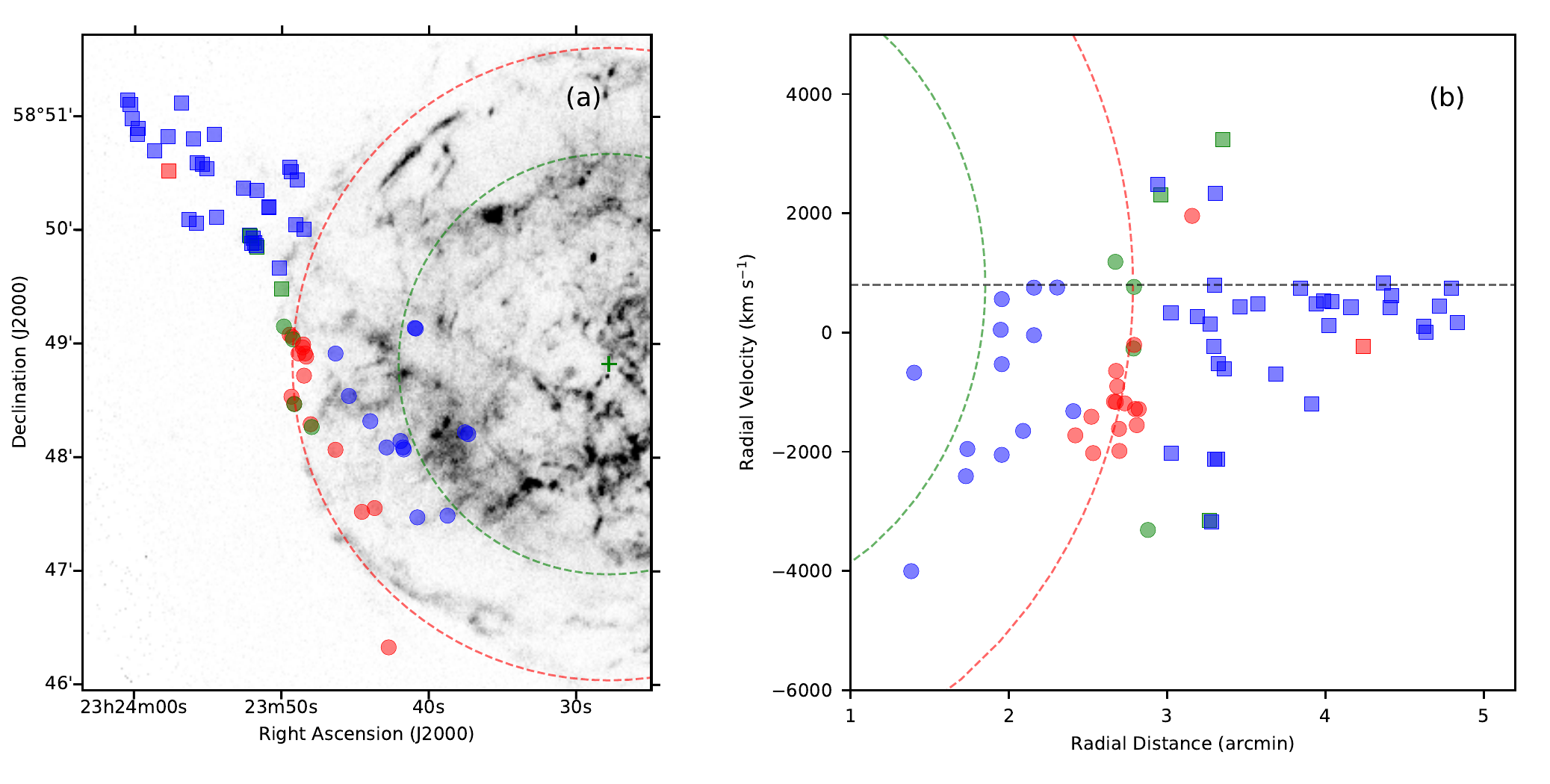}
\caption{
Spectral properties of the 67 ejecta knots shown in 
(a) (RA, Dec) and (b) (radial distance, radial velocity) planes.
The spectral properties are color-coded: 
S-rich knots in blue and Fe-rich knots in red.
The \hei\ knots are shown in green.
The background in (a) is a {\it Chandra} X-ray 4.2--6.4 keV image produced from the data
taken in 2013--2017. 
The dashed lines and the green cross symbol are the same as in Figure~\ref{fig-flxrat1}. 
The gray dashed line in the right frame indicates the systematic velocity centroid of the remnant,  
$v_{r}=+800~\kms$ \citep{ree95,del10,ise10,mil13}. 
} 
\label{fig-pvd}
\end{figure*}

\begin{figure*}
%\hspace{-0.4in}
\includegraphics[width=1.0\textwidth]{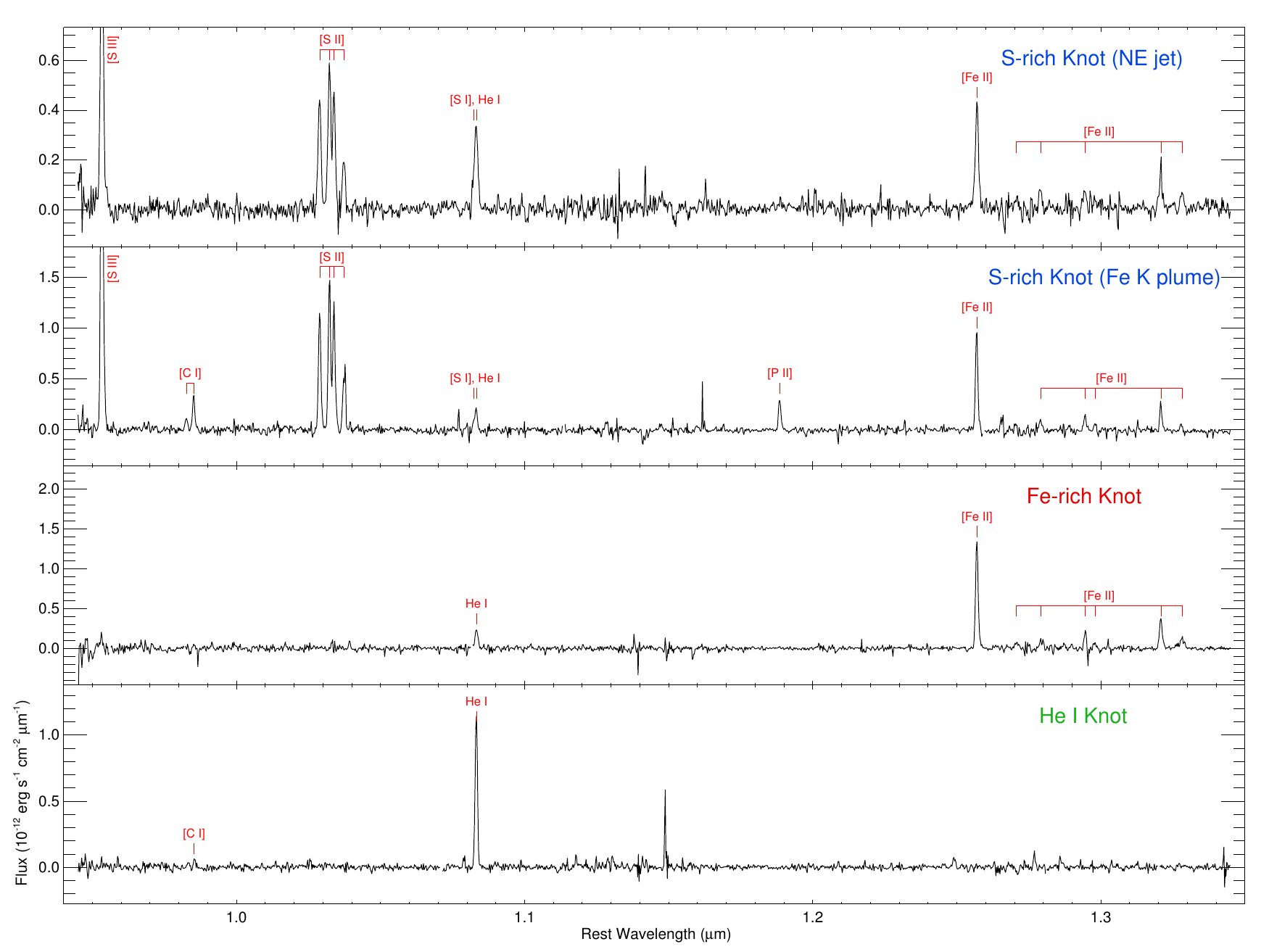}
\caption{
    Sample 1-D spectra of three ejecta knot groups:
    S-rich knots (2-06-1, 1-18-1),
    Fe-rich knots (1-16-1), and \hei\ knots (2-15-2). 
    For the S-rich knots, separate sample spectra are shown for the NE jet and Fe K plume areas because  
    the intensities of the \pii\ 1.189 \micron\ line differ significantly between the two regions. 
    The positions of the knots are marked in Figure~\ref{fig-slitpos}.
    The spiky features in the spectra
    are the residuals of bright OH airglow emission lines.
} \label{fig-1dspec}
\end{figure*}

In summary, we have classified 
the 67 knots into three groups: S-rich (44), Fe-rich (16), and \hei\ knots (7). 
Figure~\ref{fig-pvd} shows which group each individual observed knots belong to. 
We see that the majority (31/35) of the NE jet knots are S-rich knots.
Only one knot is classified as Fe-rich, which has R([S II]/[Fe II])(=1.8) close to the
threshold dividing S-rich and Fe-rich knots.
For comparison, the majority (26/32) of the Fe K plume area knots are S-rich 
or Fe-rich with clear spatial separation between them: 
the S-rich knots are near the main ejecta ring within the SNR, while
the Fe-rich knots are aligned in the north-south direction along the boundary of the SNR.
Figure~\ref{fig-pvd} also shows that 
those Fe-rich knots appear to be clustered
in the position-velocity diagram, implying that they were ejected in a narrow cone. 
We will explore the kinematic properties of the Fe-rich knots in \S~\ref{sec-dis-se}.
The \hei\ knots are detected along the forward shock front, from the southern base of the NE
jet to the Fe K plume area. Most of the \hei\ knots have large radial velocities, so 
they are different from the circumstellar 
knots bright in \hei\ 1.083~\micron\ line \citep[see][]{lee17}. 
We will explore their origin in \S~\ref{sec-dis-se}.  
Table \ref{tab-group} summarizes the properties of the three groups, and   
Figure \ref{fig-1dspec} shows the representative sample spectra of individual groups.

\subsubsection{Spectral and Physical Properties of Three Knot Groups} \label{sec-res-cla-phy}

In this section, we examine the basic spectral properties of three knot groups
and compare the densities of their line-emitting regions, 
looking for differences other than the bright line ratios.

\begin{figure*}
\begin{center}
\hspace{-0.6in}
\includegraphics[width=0.9\textwidth]{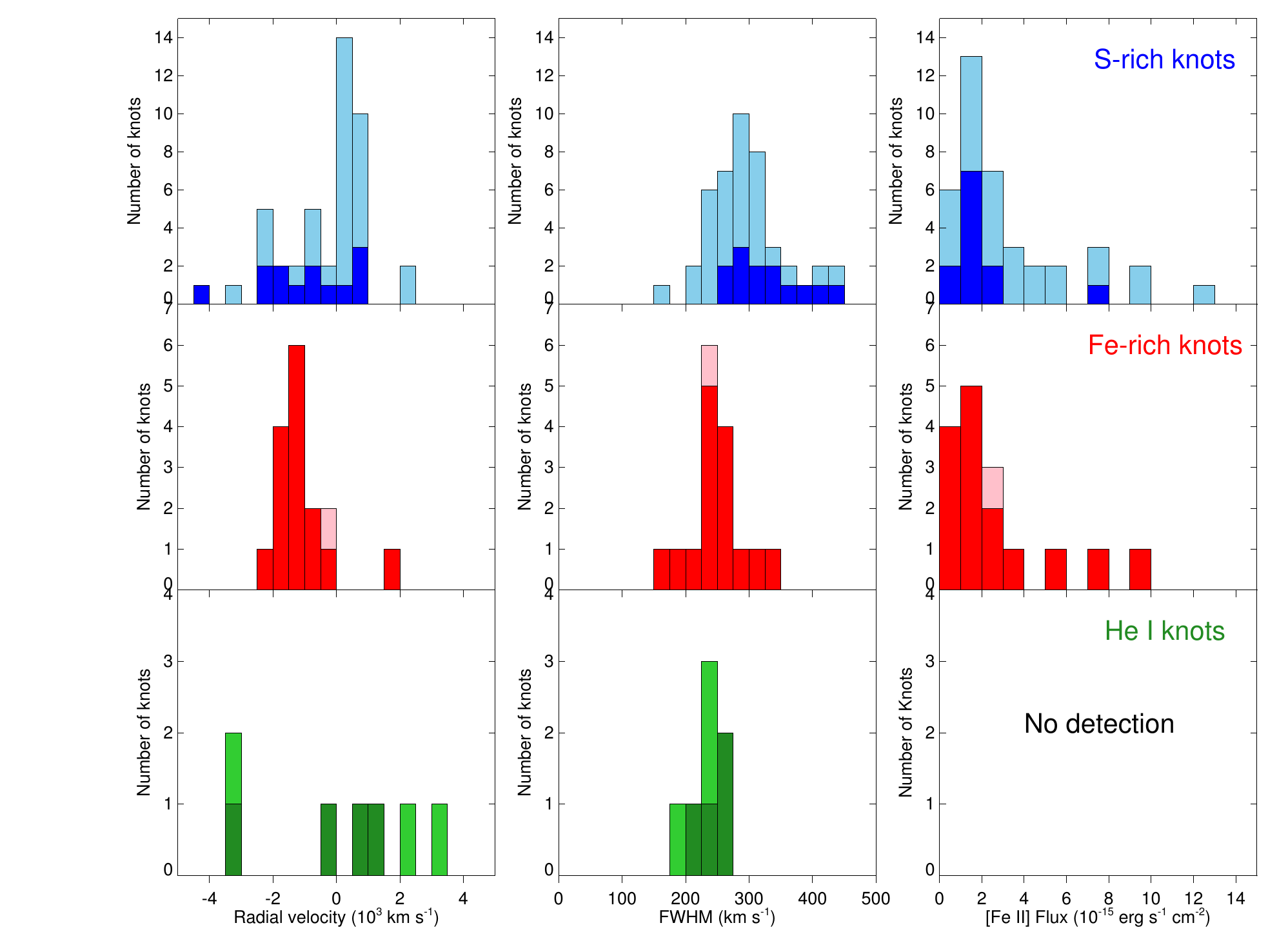}
\caption{
Distribution of the spectral parameters of the knots in three knot groups.
The columns represent different parameters: radial velocity, line width, and  
\feii\ 1.257 \micron\ flux. The rows represent different knot groups.
In each frame, the light-colored histogram represents the knots in the NE jet area and 
the dark-colored histogram represents the knots in the Fe K plume area.  
} \label{fig-histogram}
\end{center}
\end{figure*}

Figure \ref{fig-histogram} compares the distributions of the radial velocities and line widths of three knot groups. 
The radial velocities of the S-rich knots strongly peaks near 0~\kms, which 
is mostly due to the knots in the NE jet. 
The radial velocities of the knots in the NE jet are mostly confined to a narrow range 
of $v_{\rm r} = 0$ to $+800~\kms$, but there are also knots 
with high positive and high negative velocities, e.g., 
the radial velocities of the 
knots in the southernmost stream are $v_{\rm r} = -2100$~\kms\ to $-520$~\kms.
This is consistent with the previous result from optical studies that  
the bright knots in the jet streams lie close to the plane of the sky 
although the knots in the NE jet region including those in the jet base area in general
encompasses a broad range of radial velocities \citep{fes96,fes01,mil13}.
For comparison, the S-rich and Fe-rich knots in the Fe K plume area are 
mostly blueshifted. As already mentioned in \S~\ref{sec-res-cla-cla},  
the Fe-rich knots are mostly located in the Fe K plume area and their radial velocities are 
confined to a narrow range ($-2000$ to $-$200~\kms).
On the other hand, the radial velocities of the \hei\ knots 
are spread over a broad velocity range, from $-3500$ \kms\  to +3500 \kms. 
In line width, the S-rich knots have a broad distribution centered at around +290~\kms, while  
the Fe-rich and He-rich knots have a relatively narrow distribution centered at approximately +250~\kms. 
The velocity widths may be considered as a characteristic shock speed for the knots. 
But as we will see in \S~\ref{sec-res-opt}, the ejecta knots appear to have 
complex structures with subknots, 
so the shock speed could be substantially lower than this.
Figure \ref{fig-histogram} also compares the distribution of \feii\ 1.257 \micron\ line fluxes
among the three knot groups. As we already mentioned in \S~\ref{sec-res-cla-cla}, 
the S-rich and Fe-rich knots have a similar distribution of \feii\ 1.257 \micron\ line fluxes, while
the \hei\ knots have \feii\ 1.257 \micron\ line undetected with 
an upper limit mostly a few times $10^{-16}$~erg s$^{-1}$ cm$^{2}$ (see Figure \ref{fig-flxrat1}).
Note that \feii\ 1.257 \micron\ line flux in the table is the flux within the slit. 
The fluxes of the knots in the deep \feii\ image 
may be found in \cite{koo18}. 

\begin{figure*}
\vspace{0.5truecm}
\begin{center}
\includegraphics[width=0.9\textwidth]{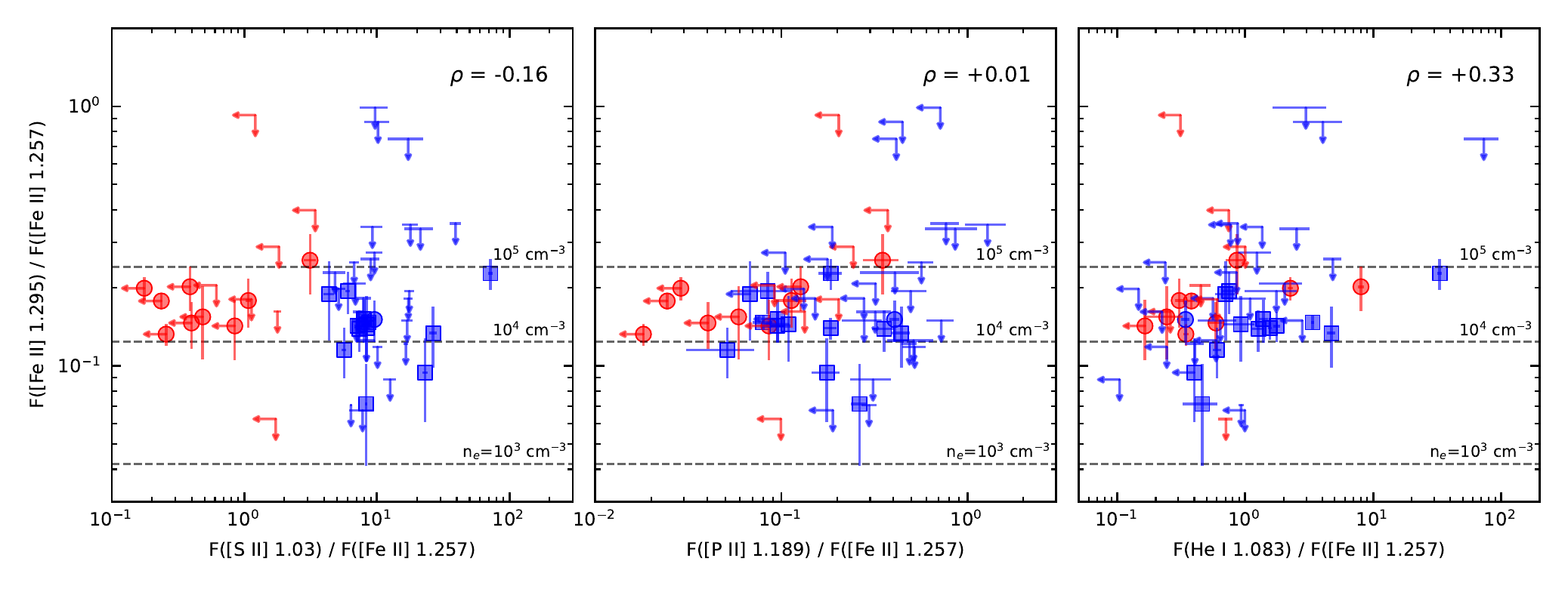}
\caption{
$F$(\feii\ 1.295)/$F$(\feii\ 1.257) versus 
\siioverfeii, \piioverfeii,  and \heioverfeii.
The knot symbols and colors are the same as in Figure \ref{fig-class}.
The horizontal lines mark the $F$(\feii\ 1.295)/$F$(\feii\ 1.257) ratios of  
ionized gas of $n_e=10^3$, $10^4$, and $10^5$~cm$^{-3}$ 
at $10^4$~K in collisional equilibrium. 
The parameter $\rho$ in each frame is the correlation coefficient between two ratios. 
} \label{fig-feii}
\end{center}
\end{figure*}
One of the physical parameters of the knots that can be
easily obtained is electron density.  In the J and H bands, 
there are several \feii\ lines originating from levels with similar excitation energies,
whose ratios depend mainly on electron densities
and can be used as a density tracer \citep[e.g., see][]{koo16,lee17}. 
In Figure~\ref{fig-feii}, we plot $F$(\feii\ 1.295)/$F$(\feii\ 1.257) of the knots, which 
is sensitive to electron density between $10^{3}$~cm$^{-3}$ and $10^{5}$~cm$^{-3}$.
The $x$-axes of the frames in the figure are 
\sii\ 1.03~\micron, \pii\ 1.189~\micron, and \hei\ 1.083~\micron\ fluxes
normalized by \feii\ 1.257~\micron\ flux.
For about two thirds of the knots,  
\feii\ 1.295 \micron\ line has not been detected, so their upper limits are plotted.   
For the knots with \feii\ 1.295 \micron\ line detected, 
the electron densities inferred from their \feii\ line ratios are mostly in the range of 
(1--4)$\times 10^{4}$~cm$^{-3}$, with the mean of 1.8$\times 10^{4}$~cm$^{-3}$.
The S-rich knots have a slightly lower density than the Fe-rich knots,
i.e, 1.5$\times 10^{4}$~cm$^{-3}$ vs 2.5$\times 10^{4}$~cm$^{-3}$.
Note that this is the density of shock-compressed, line-emitting region  
of the knots. The initial density of a knot before being shocked 
should be much lower (e.g., $\sim 10^2$~cm$^{-3}$; see Figure S3 of 
\citealt{koo13}). 
There is no clear association between the line ratios of the bright lines 
and $F$(\feii\ 1.295)/$F$(\feii\ 1.257), 
except for a moderate positive correlation ($\rho=+0.33$) observed for \heioverfeii, which 
could be possibly because the  \hei\ 1.083~\micron\ line emissivity increases with density due to the 
contribution from the collisional excitation 
from the lower level ($2 {^3}$S) \citep[e.g., see][]{koo23}.
For a given $F$(\feii\ 1.295)/$F$(\feii\ 1.257) ratio, the 
\siioverfeii\ of S-rich and Fe-rich knots differs by more than an order of magnitude. 
This suggests that the density of the shocked gas is 
not the primary factor causing the division between the two groups of ejecta knots.

\subsection{Optical Counterparts} \label{sec-res-opt}

Numerous dense ejecta knots have been identified around 
the Cas A SNR in previous optical studies, and 
it will be interesting to check if our NIR ejecta knots have optical counter parts. 
We use the optical ejecta knot catalog of 
\citet{ham08}, who identified 1825 compact optical knots
that lie beyond a radial distance of $\sim 2\arcmin$ from the explosion center
in {\it Hubble} ACS/WFC images taken with three different broadband filters 
(i.e., F652W, F775W, F850LP) in 2004.
They were able to classify the optical knots into three groups based on 
their flux ratios in the three filters:
(1) \nii\ knots dominated by [\ion{N}{2}] $\lambda\lambda$6548, 6583 emissions, 
(2) \oii\ knots dominated by [\ion{O}{2}] $\lambda\lambda$7319, 7330 emissions, and (3) 
Fast-Moving Knot (FMK)-like knots displaying filter flux ratios suggestive of 
\sii, \oii, and [\ion{Ar}{3}] $\lambda$7135
emission line strengths similar to the FMKs found in the remnant's 
main ejecta shell (see also \citealt{fes06}).
Of the 1825 knots identified by them, 
444 were \nii\ knots, 192 were \oii\ knots, and 1189 were FMK-like knots.
Their spatial distribution are distinct; 
\nii\ knots are arranged in a broad shell around the remnant, 
\oii\ knots are clustered around the base of jets,   
and FMK-like knots are mainly confined to NE and SW jet areas (see also \citealt{fes16}).
We have compared the spatial positions of our 67 NIR ejecta knots with
those of the optical outer knots shifted to 2017 epoch assuming free expansion.
Considering the seeing and slit width of our observations,
we have searched their optical counterparts within 0\farcs5 radius.
Note that 12 out of the 67 ejecta knots are located inside $r=2\farcm4$
where \citet{ham08} did not search optical knots,
so we can check the counterparts only for 55 NIR knots.

\begin{figure*}[t]
\includegraphics[width=\textwidth]{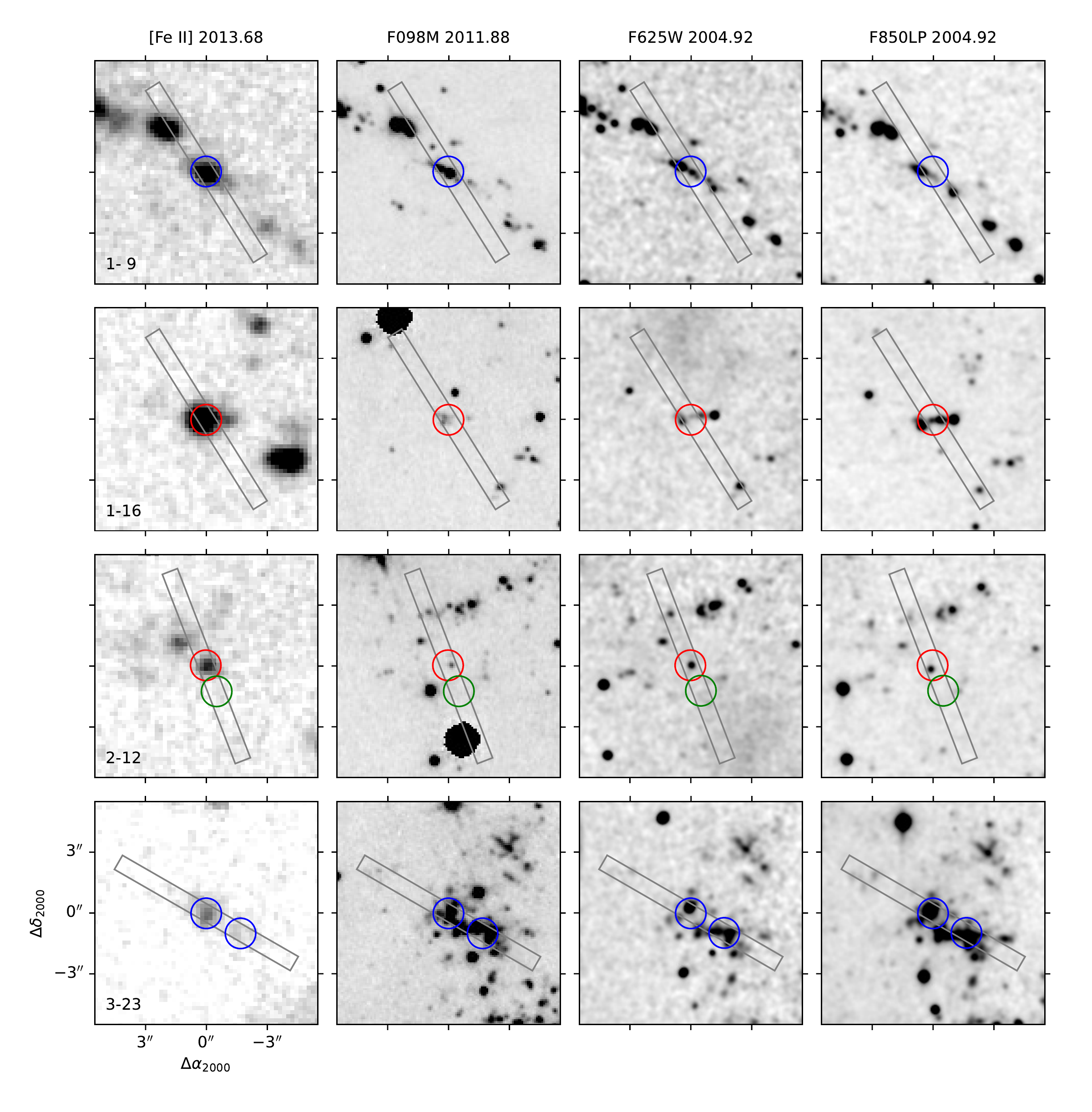}
\caption{
Optical counterparts of the sample of NIR knots detected 
in slits 1-9, 1-16, 2-12, and 3-23.
For each slit area, four images are shown: 
the deep \feii\ image and three \hst\ images obtained with 
F098M, F625W, and F850LP filters. 
North is up, and east is to the left.
In each frame, the gray rectangle represents the location of the 10\arcsec-long MOS slit, and 
the circles mark the knot locations, the colors of which represent the knot groups: 
       S-rich knots in blue, Fe-rich knots in red, and \hei\ knots in green.
Note that the positions of the slits and the circles are shifted to the epochs of the 
background images, which are given at the top of individual columns.
} \label{fig-counter}
\end{figure*}

\begin{figure*}
\begin{center}
\includegraphics[width=0.9\textwidth]{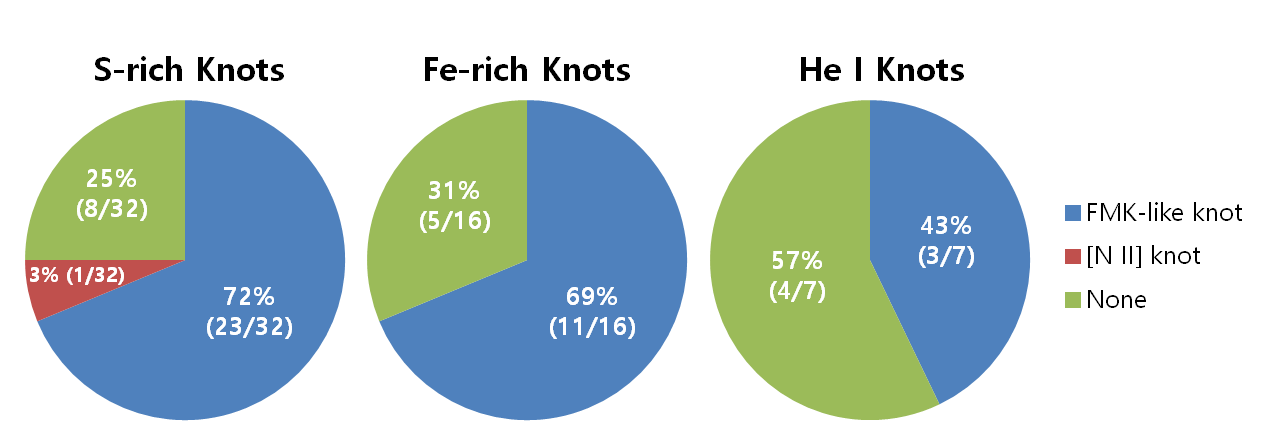}
\caption{
    Pie charts showing the optical counterparts of 55 NIR ejecta knots.
    The red and blue slices represent the percentages of the NIR knots with
    optical \nii\ and FMK-like knots as counterparts, respectively, while
    the green slice represents the percentage of the NIR knots without optical counterparts.
    Note that the statistics of S-rich knots includes only 32 knots out of 44. 
    The rest twelve S-rich knots are located in the inner region of the Cas A SNR 
    where the characteristics of the optical knots were not investigated by \citet{ham08}. 
} \label{fig-opt}
\end{center}
\end{figure*}

We found that 38 out of the 55 NIR knots have optical counterparts. 
All of them have FMK-like knots as a counterpart except one that
appears to have an \nii\ knot as a counterpart. No NIR knots have \oii\ knots as a counterpart. 
It is not unusual that the NIR knots that are identified from 
the ground-based, deep \feii\ image are resolved into multiple knots in high resolution {\it HST} images, implying that the knots can have a complicated structure consisting of subknots. 
Figure \ref{fig-counter} shows optical counterparts of the sample of the observed NIR knots. Some NIR knots appear as a single, very compact optical knot, but they often have a more complex structure that is resolved into smaller subcomponents. In some cases, an NIR knot is found to be a part of a larger, more extended structure in optical. We further note that optical knots tend to cluster together with other knots of the same type. However, there are rare cases where different types of optical knots are found in close proximity to each other, especially in the main shell region. While these can complicate the identification of optical counterparts, we do not think the overall trend will be affected.
Figure~\ref{fig-opt} shows the pie charts for the optical counter parts
of the three groups of the NIR knots.
About three quarters of S-rich and Fe-rich knots
have optical counterparts, while about half of \hei\ knots has optical counterparts. 
It is worth noticing that
one quarter of S-rich knots has no optical counterpart, although, 
since the F850LP filter is sensitive to
\sii\ 1.03~\micron\ and \siii\ 0.907, 0.953~\micron\ lines,
the S-rich knots emitting strong \sii\ and \siii\ lines
should be detected at least in the deep F850LP image.
These S-rich knots not detected in the deep optical image
may have appeared after 2004 when the {\it HST} images were taken:
It has been found that the knots in the NE jet and Fe K plume 
areas reveal substantial emission variations
(``flickering'') in an interval as short as nine months (\citealt{fes11}; see also Figure~\ref{fig-counter}).
Indeed, we were able to confirm that 
several knots (e.g., Knots 2, 7, 15) were not present in the 2004 {\it HST} image but 
appeared in a subsequent 2010 {\it HST} image.  

We also found that 70\% of the Fe-rich knots have FMK-like knots as their optical counterparts.
This is inconsistent with our results that  
most Fe-rich knots do not have \sii\ 1.03~\micron\ and \siii\ 0.953~\micron\ lines.
One possibility is that
the flux detected in the F850LP image is not due to \sii\ or \siii\ lines
but \feii\ lines. 
There are several bright \feii\ lines, e.g.,  
8619 \AA, 8894 \AA, 9229 \AA\ \citep{den86,hur96,koo16},   % in Air wavelength
in the wavelength band (8000--10{,}000 \AA) of the 
F850LP filter, while no or little bright \feii\ lines in the wavebands of the F652W and F775W filters.  
For example, the expected \feii\ $\lambda$8619/\feii$1.257\mu$m ratio ranges from 
0.08--0.49 for electron densities $10^3$--$10^5$ cm$^{-3}$ \citep{koo16}. 
Hence, \feii\ lines can make a substantial contribution to the F850LP filter flux, and,  
together with the high interstellar extinction toward Cas A,
the Fe-rich knots in out sample could have been classified as 
FMK-like knots in optical band.
Note that the majority of the optical knots in the 
catalog of \cite{ham08} are very faint
with the F850LP flux less than $5\times 10^{-17}$~erg cm$^{-2}$ s$^{-1}$,
which is more than two orders of magnitude smaller than the \feii\ 1.257 \micron\ fluxes 
of Fe-rich knots (Figure~\ref{fig-fxvsfx}).
Another possibility is that the knots have    
\sii\ and/or \siii\ lines but the lines are `weak', i.e., \siioverfeii $<3.7$, 
so they have been classified as Fe rich knots in this work, but 
they were detected in the deep F850LP image and classified as FMK-like knots. 
Indeed, at least three of the Fe rich knots with FMK-like knot counterparts 
do have \sii\ lines (see Table~\ref{tab-feknots} in \S~\ref{sec-dis-se}).
Yet another possibility is a chance coincidence. The area where the 
Fe-rich knots have been detected are crowded with the 
FMK-like knots (e.g., see Figure 1 of \citealt{ham08}), 
so that we cannot rule out the possibility that a 
FMK-like knot accidentally falls within the circle of $0\farcs5$ radius. 
But, since the distances between the matched knots are usually an order of magnitude 
smaller the median distance between the optical knots around Fe-rich knots (i.e., a few arcseconds), 
most of the matches are probably genuine.

Lastly, it is interesting that about half of the \hei\ knots has 
FMK-like knots as their optical counter parts. 
This may indicate that \hei\ knots are just S-rich knots 
with \sii, \siii, and \feii\ lines below our detection limit.
But, since most of the \hei\ knots are also located 
in the area crowded with FMK-like knots, a chance coincidence is 
possible. We will explore the nature of \hei\ knots in \S~\ref{sec-dis-he}.

\section{Discussion} \label{sec-dis}

\subsection{S-rich Ejecta Knots in the NE Jet} \label{sec-dis-ne}

A most striking morphological feature in Cas A 
known from early optical studies is 
the NE jet structure \citep{van70,fes96}. 
It is composed of the streams of bright knots 
confined into a narrow cone well outside
the northern SN shell that appears to have 
punched through the main ejecta shell.  
The jet structure extends to $\sim 3'$ out beyond the main shell 
implying the ejection velocities upto 16,000 \kms, 
which is three times faster than the bulk of the O- and S-rich ejecta in the 
main ejecta shell \citep{kamper76,fes96,fes01,fes16}. 
Later the SW ``counterjet'' composed of numerous optical knots 
with comparable maximum expansion velocities  
was detected on the opposite side \citep{fes01,mil13}.
The origin of the bipolar NE jet--SW counterjet structure is not fully understood
although it must have been generated very near the explosion center 
during the first seconds of the explosion (see below).
A critical information for the origin of the jet comes from its chemical composition.
Previous optical studies showed that the NE jet knots exhibit   
strong S, Ca, and Ar emission lines with comparatively weak
oxygen lines \citep{van71,fes96,fes01,ham08,mil13,fes16}. 
The knots lying farther out and possessing the highest expansion velocities 
show no detectable emission lines other than those of \sii\ \citep{fes16}.
The knots near the jet base, on the other hand, show O and N emission too \citep{fes96}.
The jet structure has been observed in X-ray and IR too. 
In X-ray, the jet material shows strong Si, S, Ar, and Ca lines  
but not particularly strong Fe K line \citep{hug00,hwa04,vink04}.
Fe K emission has been found to be strong in other areas of Cas A, 
particulary in the eastern area \cite[][see \S~\ref{sec-dis-se}]{hwa03,hwa12,picquenot21}. 
These X-ray studies showed that 
the elemental composition of the jet, enriched in the intermediate-mass elements with 
a limited amount of Fe, is similar to that of oxygen or incomplete 
Si burning (see also \citealt{ikeda22}).
In IR, the ring-like jet base lifted from the main ejecta shell is pronounced 
in Ar emission \citep{del10}.
These findings suggest that the jet originated in the deep interior of  
the progenitor star where Si, S, Ar, and Ca are nucleosynthesized and penetrated 
through the outer stellar envelope.

\begin{figure*}
\begin{center}
\includegraphics[width=0.9\textwidth]{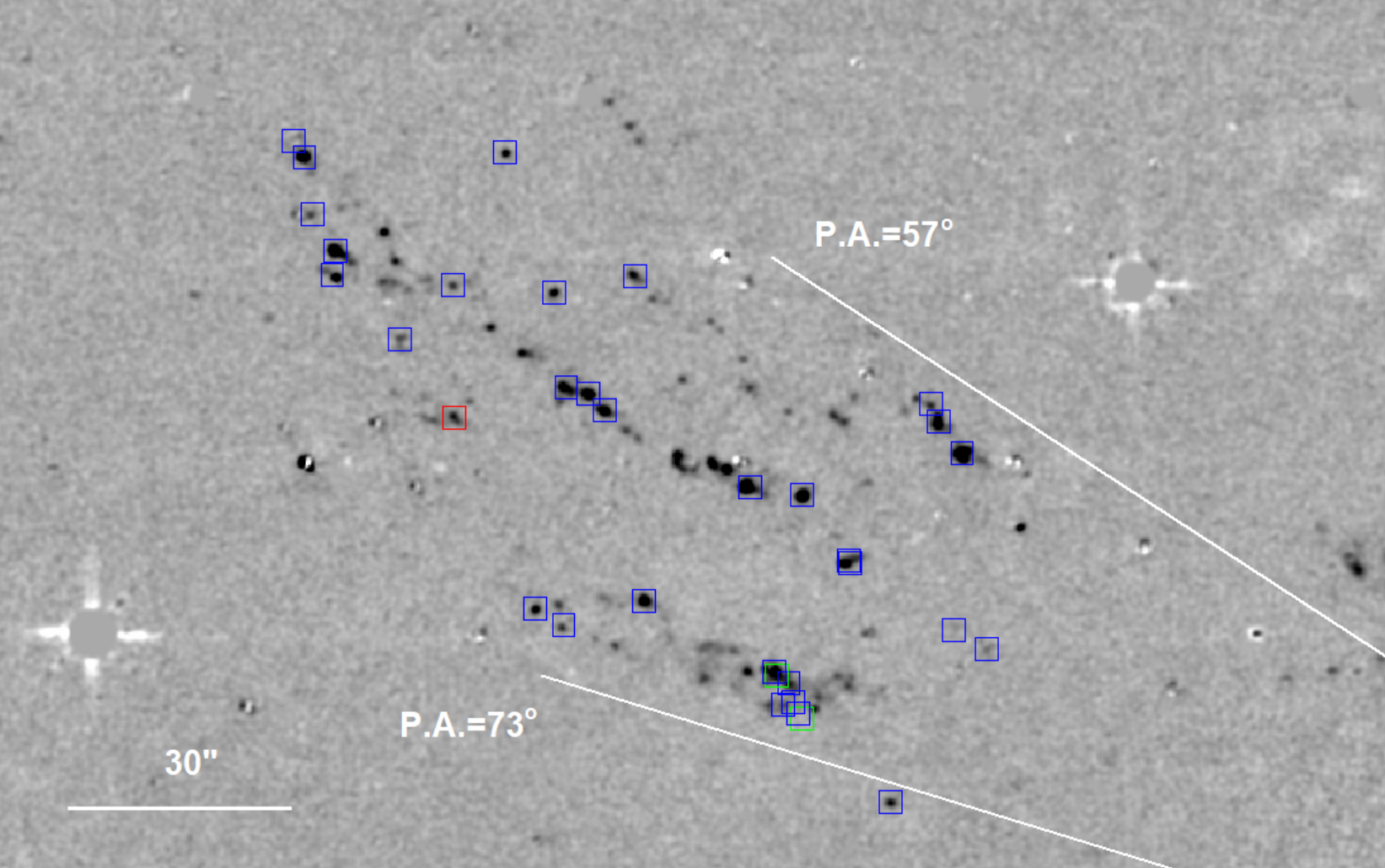}
\caption{
	Detail of the NE jet area. 
	The background is the deep \feii\ image. North is up, and east is to the left.
	The knots where NIR spectra have been obtained in this work are marked by 
	 squares, the colors of which represent the knot groups: 
       S-rich knots in blue, Fe-rich knots in red, and \hei\ knots in green.
} \label{fig-ne_detail}
\end{center}
\end{figure*} 

\begin{figure*}
\begin{center}
\includegraphics[width=0.6\textwidth]{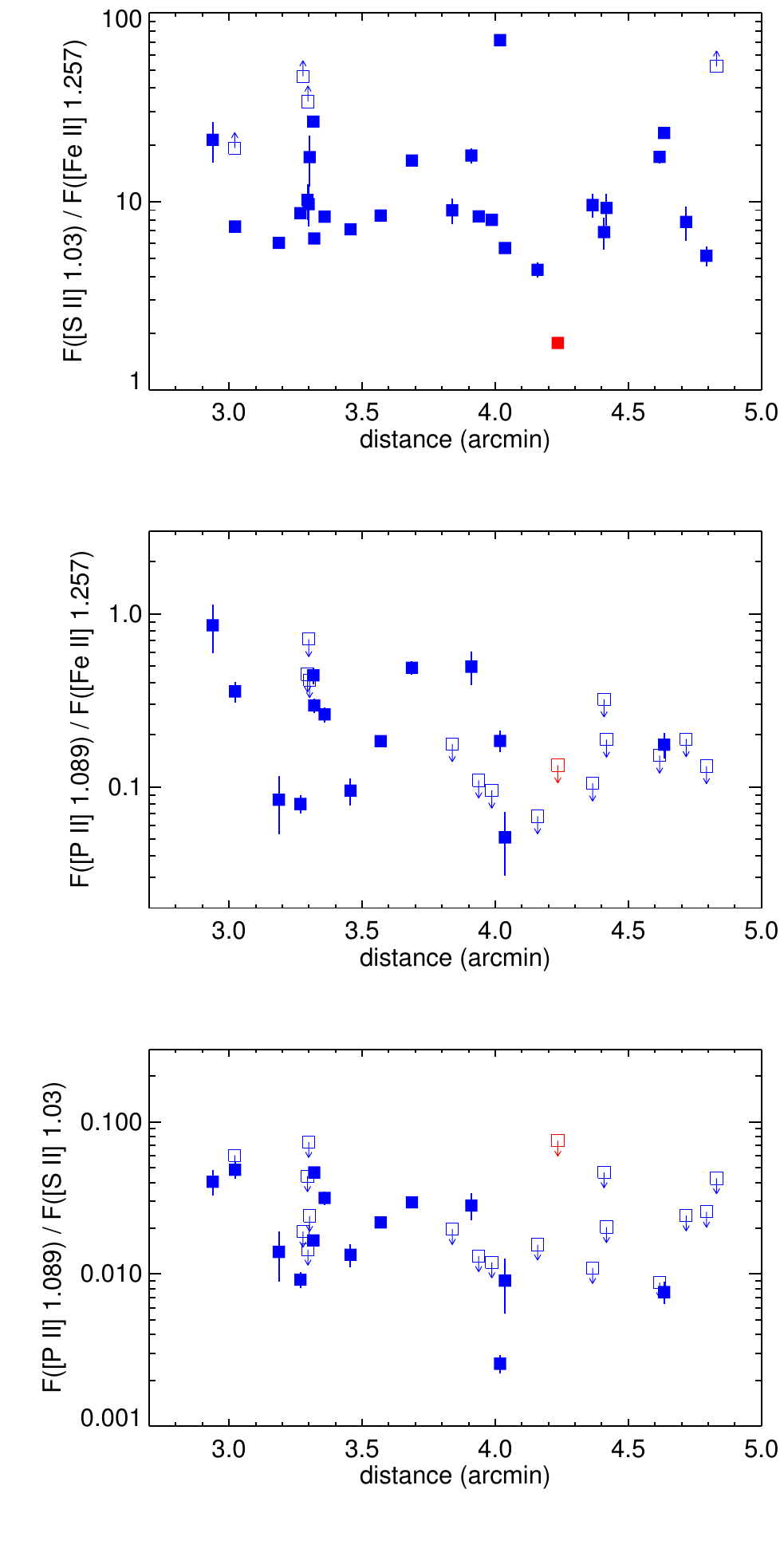}
\caption{
Line ratios of the NE jet knots as the function of distance from the explosion center.  
S-rich and Fe-rich knots are denoted by blue and red squares, respectively, while
empty symbols indicate upper limits.  
} \label{fig-ne_ratio}
\end{center}
\end{figure*}

Figure \ref{fig-ne_detail} gives a detailed view of the NE jet 
structure in \feii\ 1.644~\micron\ emission.
In the image, the NE jet appears as streams of bright knots 
confined to a fanlike structure with an opening angle of 10$\degr$ 
viewed at the explosion center.  
The central stream at P.A.$\simeq 64\degr$ is most prominent, 
extending to $4\farcm7$($= 4.6$ pc) from the explosion center. 
In the northern area above the central stream, there are many bright knots 
scattered over the area that appear to be aligned along several stream lines, while, 
in the southern area below the central stream,  
there is a prominent but fuzzy stream of ejecta material possessing a finite width 
at P.A.$\simeq70\degr$, and not many bright knots are in between.
This jet morphology is very similar to what we have already seen in  
\sii\ $\lambda\lambda$\ 6716, 6731 emission \citep{fes96}.

In Figure\ref{fig-ne_detail}, we marked the ejecta knots observed in this work using 
different colors for S-rich, Fe-rich, and \hei\ knots. 
We see that the NE jet knots are mostly S-rich (see also Figure \ref{fig-pvd}). 
We have not detected an Fe-rich knot exhibiting only \feii\ lines.
There is one knot classified as a Fe-rich knot located 
between the central and the southern streams, 
but it exhibits \sii\ lines of moderate strength (\siioverfeii=1.8).
These results are consistent with the previous results from optical and X-ray studies
that the jet material shows strong S lines but not particularly strong Fe lines.
In optical bands, the NE jet knots show strong O, S, Ar lines 
like the other S-rich knots and are not clearly distinguished  
from the S-rich knots (FMKs) in the main ejecta shell,  although, for example, 
a few outermost knots show stronger than usual 
[\ion{Ca}{2}] $\lambda\lambda$7291, 7324 lines that could be due to   
abundance difference \citep{fes96}. 
In NIR, however, the NE jet knots are distinguished from  
the S-rich knots in the main ejecta shell in \pii\ emission: 
The majority of the dense knots in the NE jet show 
weak or no \pii\ lines whereas the S-rich knots in the main ejecta shell 
show strong \pii\ lines. This is shown in Figure \ref{fig-ne_ratio},
where the line ratios of the NE jet knots 
are plotted as a function of distances from the explosion center (see also Figure~\ref{fig-flxrat1}).
The figure shows that \siioverfeii\ remains roughly constant along the jet, 
whereas \piioverfeii\ and \piioversii\ appear to decrease systematically along the jet.
Note that \pii\ line has not been detected in the majority of the 
knots near the tip of the jet. 
For comparison, the 
three ratios for the S-rich knots in the main ejecta shell are 
(\siioverfeii, \piioverfeii, \piioversii)=($14.2\pm 13.5$, $1.46\pm1.31$, $0.117\pm 0.083$) 
\citep[see Figure 2 of ][]{koo13}.
 
The simplest interpretation of the low \pii/\feii\ and \pii/\sii\ ratios 
of the NE jet knots is that  
the amount of phosphorus present there is lower compared to the knots in the main shell.
Although the knots in the NE jet and the main shell have 
different shock environments, the decrease in these ratios  
from the main shell to the tip of the NE jet, while 
the \sii/\feii\ ratio remains constant, suggests that 
the phosphorus abundance is probably the primary factor affecting the line ratios.
Therefore, we consider the weakness of the \pii\ line in the NE jet knots as  
direct evidence that  
the NE jet originated from a layer with a high sulfur and low phosphorus content.
And, since $^{15}$P is mainly produced in carbon and neon burning layers \citep{arnett96,woosley02}, 
it implies that the NE jet was ejected from 
a region below the explosive Ne burning layer, i.e., O burning or incomplete Si burning layers.
This is consistent with the result of previous X-ray studies, which showed that 
the elemental composition of diffuse ejecta material in the NE jet is 
similar to that of oxygen and incomplete burning \citep{vink04,ikeda22}. 

We now discuss the implication of our results on the physical origin of the NE jet. 
The origin of the NE jet and its role in the SN explosion 
have been subject to study since its discovery in 1960s.
\cite{min68} suggested that the jet structure (or the `flare' in his paper) 
could be a surviving part of a spherically symmetric shell due to non-spherically 
distrbution of ambient interstellar medium \citep[see also][]{fes96}.
But it is now rather well established that 
the NE jet originated from a Si-S-Ar rich layer deep inside the progenitor 
and penetrated through the outer 
N and He-rich envelope \citep{fes01,fes06,lam06,mil13}.
The chemical composition of the jet material 
and the spatial correlation between the jet and the disrupted main ejecta shell 
are strong evidence for that.
On its launching mechanism, there have been several theoretical propositions  
in relation to the core-collapse SN explosion models. 
In the classical `jet-induced' explosion model, where 
magnetic field anchored to the collapsing core 
is amplified during the collapse and drives the SN explosion, 
a magnetocentrifugal jet can be ejected along the rotation axis \citep{kho99,whe02,akiyama03}.
Recent observational studies, however, 
suggest that the properties of the NE jet are not consistent with 
the jet-induced explosion model; 
there is no `pure' Fe ejecta in the jet \citep{ikeda22},   
the opening angle is large and the kinetic energy is insufficient, and 
the kick direction of the neutron star kick is not along but perpendicular to the jet axis  
(\citealt{fes01,hwa04,lam06,mil13,fes16}; cf. \citealt{soker22}).
Our results are also in line with previous results and do not support the jet-scenario.
We have not found dense Fe-rich knots around the tip of the NE jet either in this study. 
The opening angle of the jet in Figure~\ref{fig-ne_detail} is small ($\simlt 15\degr$), 
but their total kinetic energy will be less than the 
previous estimate of $1\times 10^{50}$~erg of the optical knots in the NE jet \citep{fes16}. 
Hence, our results also do not 
support the jet-induced explosion model.

In the neutrino-driven explosion model, which is the modern paradigm of 
the core-collapse SN explosion, a narrow high-velocity ejecta  
can be generated either during the explosion or  
after the explosion by the newly-formed neutron star. 
During the explosion, high-speed jet-like fingers can be produced 
from the core by neutrino convection bubbles \citep[e.g.,][]{bur95,jan96}. 
In particular, the interface between Si and O composition is unstable
seeded by the flow-structures resulting from neutrino-driven convection, 
leading to a fragmentation of this shell into metal-rich clumps \citep{kifonidis03}.
\cite{orlando16} showed that the Si-rich, jet-like features such as the NE jet in Cas A could be 
reproduced by introducing overdense knots moving faster than the surrounding 
ejecta just outside the Fe core. 
But a recent three-dimensional simulation for modeling the Cas A SNR as a remnant of 
a neutrino-driven SN, starting from the core-collapse to the fully developed remnant,  
could not reproduce a structure similar to the NE jet \citep{orlando21}. 
Another possibility is that hydrodynamic jet from a newly-formed neutron star.
In a standard neutrino-driven explosion, a fraction of ejecta falls back to 
the newly-formed neutron star to form a disk where 
a MHD jet can be launched \citep{bur05,burrows07,burrows21,janka22}. 
 But, as far as we are aware, there is little theoretical prediction
that can be compared with observation for such an MHD jet. 
So for the neutrino-driven explosion model, 
it is still an open question if the NE jet can be explained by this model.

\subsection{Fe Ejecta Knots in the Fe K plume area} \label{sec-dis-se}

\begin{figure*}
\center
\includegraphics[width=0.9\textwidth]{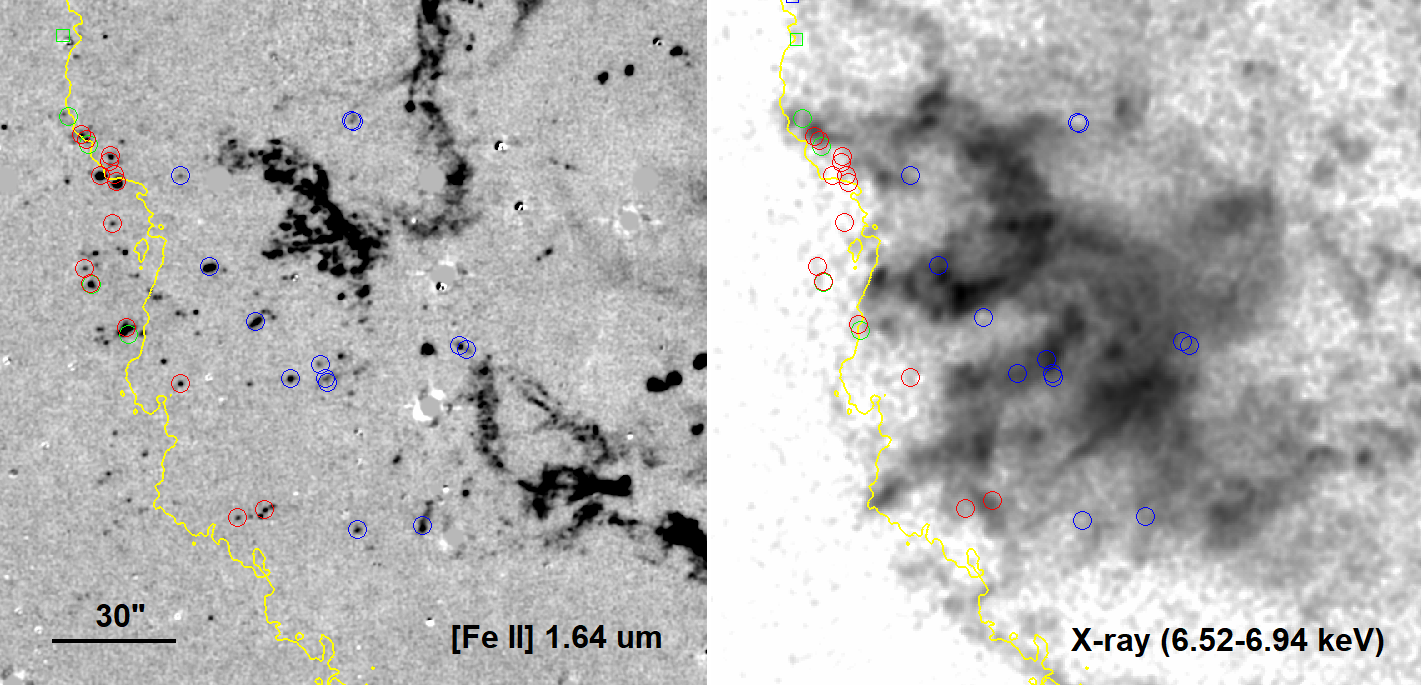}
\caption{
Detail of the Fe K plume area in the deep \feii\ and 
{\it Chandra} X-ray Fe K emission images. North is up, and east is to the left.
The knots where NIR spectra have been obtained in this work are marked 
by circles (P.A.$\ge78^\circ$) or squares (P.A.$<78^\circ$), 
the colors of which represent the knot groups: 
       S-rich knots in blue, Fe-rich knots in red, and \hei\ knots in green.
Note that the positions of the 
knot markers are shifted to the epochs of the background images, 
i.e., 2013 Sept. for the deep \feii\ image 
and 2004 Apr. for the {\it Chandra} image. 
The yellow contour marks an approximate boundary of the SNR 
%in radio corresponding to the intensity level of 0.3 mJy beam$^{-1}$ 
in radio in 2003 \citep{delaney04}, and it is left as a reference point.
} \label{fig-se_detail}
\end{figure*}
We discovered Fe-rich knots in the Fe K plume area (see Figure \ref{fig-se_detail}). 
This is the area where the Fe-rich plume with bright Fe K emission has been detected in X-ray 
\citep{wil02,hwa03,hwa04,hwa12,tsuchioka22}.
The ejecta material in this Fe K plume has no S or Ar, and its Fe/Si abundance 
ratio is up to an order of magnitude higher than
that expected in incomplete Si-burning layer, indicating that they originated from 
complete Si-burning process with $\alpha$-rich freezeout, 
where the burning products are almost exclusively $^{56}$Fe 
because free $\alpha$ particles are unable to reassemble to
heavier elements on a hydrodynamic time scale \citep{arnett96,woosley02}.
Recently, \cite{sato21} detected Ti and Cr in this area, 
the ratio of which to Fe is consistent with explosive complete Si burning with 
$\alpha$-rich freezeout.
Therefore, the material in the Fe K plume has likely been produced in the 
innermost, high-entropy region during the SN explosion 
where the temperature is high and density is low \citep[e.g.,][]{arnett96, thi96}.
In two-dimensional projected maps, however, the Fe K plume is located much further outside 
the main shell which is composed of ejecta synthesized from 
incomplete Si burning, e.g.,  Si, S, Ar,
which could be interpreted as a overturning of ejecta layers \citep{hug00}. 
But a three-dimensional reconstruction of the ejecta distribution   
using infrared and {\it Chandra} X-ray Doppler velocity measurements 
showed that, in the main shell, the Fe K plume occupies a ``hole'' surrounded 
by a ring structure composed of Si and Ar 
 (\citealt{del10}; see also \citealt{mil13}). 
As an explanation of the observed morphology, 
\cite{del10} proposed a model where an ejecta ``piston''    
faster than average ejecta has pushed through the Si layers 
rather than interpreting it as an overturning of the layers. 
In this model, the Si-Ar ring structure in the main shell 
represents the circumference of the piston intersecting with the reverse shock. 
Recently \cite{tsuchioka22} showed that the velocity of the 
ejecta in the outermost Fe K plume is $> 4500$~\kms,  
much higher than that of Si/O-rich ejecta, suggesting that the ejecta piston 
has been formed at the early stages of the remnant evolution or during the SN explosion,  
which is consistent with the results of numerical simulations. 
On the other hand, recent 3-D numerical simulations 
demonstrated that Ni clumps, created 
during the initial stages of the SN explosion in the innermost region by the
convective overturning in the neutrino-heating layer 
and the hydrodynamic instabilities, 
later decay to $^{56}$Fe and can push out the less dense ejecta 
to produce structures such as Fe K plume \citep{orl16, orlando21}. 
But \cite{hwa03} pointed that the morphology and 
the `ionization age' ($=n_e t$ where 
$n_e$ is electron density and $t$ is the time since the plasma was shock-heated)
of the Fe ejecta in the Fe K plume are inconsistent with a Ni bubble origin.

Our detection of Fe-rich knots in the Fe K plume area has quite significant implication.
Figure \ref{fig-se_detail} shows the distribution of IR ejecta knots in the Fe K plume area. We see a group of  Fe-rich knots aligned in the north-south direction along the forward shock front. 
They are spatially confined in the area immediately outside the boundary 
of the diffuse X-ray Fe ejecta.
Their radial velocities are in a narrow range, from $-2000$~\kms\ to $-$200~\kms\ (see Figure~\ref{fig-histogram} and Table \ref{tab-feknots}), 
so, in a position-velocity diagram, they appear to be clustered (see Figure \ref{fig-pvd}b).
The ejection angle of this ``Fe-rich ejecta knot cluster'' is 
$14\degr \pm 5\degr$ from the plane of the sky toward us.
(In Figure \ref{fig-pvd}b, we see two Fe-rich knots located far off the cluster.  
One of them is Knot 23 in the NE jet area and the other is 
Knot 67 near the southern edge of the field. The latter 
has a radial velocity +1960~\kms, which is very different from those of the Fe-rich ejecta knot cluster. 
They are not considered to belong to the cluster.)
The radial velocities of the X-ray emitting diffuse ejecta in this area measured with 
{\it Chandra} High Energy Transmission Grating are from $-2600$ to $-500$~\kms, comparable with those of
the Fe-rich knots \citep{lazendic06,rutherford13}.
On the other hand, 
the proper motion of Fe-rich knots has velocity of 6600--7600~\kms\ (Table \ref{tab-feknots}), which  
is considerably higher than those of the diffuse ejecta 
at the tip of the Fe K plume (i.e., 4500--6700~\kms, \citealt{tsuchioka22}).  
Hence, the Fe-rich knots move a little faster than the diffuse X-ray emitting Fe-rich ejecta.
The spatial and kinematic relations strongly support the 
physical association between the dense Fe-rich knots and the X-ray diffuse Fe ejecta.
And the spectral properties of the Fe-rich knots suggest
that they are also produced by explosive Si complete 
burning with $\alpha$ freezeout as the diffuse Fe ejecta.
The \hei\  1.083~\micron\ lines 
detected in the majority of Fe-rich knots might be from helium remains 
of the $\alpha$-rich freeze-out process (Table~\ref{tab-feknots}): 
In $\alpha$-rich freezeout, the burning products are 
almost exclusively $^{56}$Fe, but a considerable amount of He could remain. 
The local He mass fraction can be as high as 0.2 depending on model 
\citep{rauscher02,woosley02,thi18}.
The detection of dense Fe ejecta knots   
implies that the Fe ejecta produced in the innermost region was very inhomogeneous.     

Theoretically, dense metallic ejecta knots 
can be produced during the explosion. 
As was mentioned above, 
numerical simulations have shown that when the 
innermost ejecta expanding aspherically    
encounters a composition shell interface, it is compressed to a dense shell, which 
fragments into small, dense metallic clumps by hydrodynamic instabilities
\citep{kifonidis00,kifonidis03,hammer10,wongwathanarat17}. 
It is uncertain whether the dense knots had initially higher velocities. 
However, it is possible that these dense knots were initially comoving  
with the surrounding diffuse ejecta, but once they encounter the reverse shock, 
the dense and diffuse ejecta are decoupled, 
i.e., the diffuse Fe ejecta are decelerated more, 
so the dense Fe clumps are located ahead of the diffuse Fe ejecta.
This explanation has been proposed for dense 
Fe knots beyond the SNR boundaries detected 
in X-rays in young Type Ia SNRs \citep{wan01,tsebrenko15}. 
A caveat in this scenario is, since the radioactive decay from $^{56}$Ni to $^{56}$Fe produces
post-explosion energy, the Fe clumps are expected to expand 
unless they are optically thin to gamma-ray emission
and should be characterized by the diffuse morphology in the remnant phase 
\citep{blo01,hwa03,gabler21}. 
Indeed, in the interior of Cas A,  the bubble-like structures in S-rich ejecta 
filled with $^{44}$Ti are detected supporting this scenario \citep{mil13, gre17}.
Therefore, the dense compact Fe-rich knots discovered in Cas A need an explanation.  
A possible explanation is 
that the dense Fe-rich knots are 
not $^{56}$Fe decayed from $^{56}$Ni but $^{54}$Fe \citep{wan01}. 
But $^{54}$Fe is not a dominant element \citep[see][and below]{hwa03}.
In core-collapse SN particularly, $^{54}$Fe turns to $^{58}$Ni (stable nuclei of nickel) in
$\alpha$-rich freezeout region of complete Si-burning zone,
so it is only detected in incomplete Si-burning layer,
where S and Si abundances are comparable with
or much larger than the $^{54}$Fe and $^{56}$Ni/$^{56}$Fe abundances
\citep[e.g.,][]{thi96,thi18}.
Therefore, Fe in our Fe-rich knots is most likely $^{56}$Fe, not $^{54}$Fe.
Another possibility, which was proposed by \citet{hwa03} to explain 
the compact X-ray knots embedded in the Fe K plume, 
is that the optical depth of the knots  
is not large enough to the $\gamma$-ray from the radioactive decay from 
$^{56}$Ni to $^{56}$Fe.
Using a hydrodynamics model for ejecta evolution, 
they showed that, for a typical knot of diameter $\sim 3''$ 
and electron density $\sim 10$~cm$^{-3}$, 
the optical depth of $\gamma$ rays ($\sim 1$ MeV) 
becomes less than one about 5 days after explosion, which is 
much shorter than the $^{56}$Ni and $^{56}$Co lifetimes, i.e., 
8.8 days and 111.3 days, respectively.
We consider that the Fe-rich knots discovered in our NIR study
could be a denser and faster version of the X-ray knots.
The NIR Fe-rich knots are dense 
(e.g., $\simgt 10^2$~cm$^{-3}$; \S~\ref{sec-res-cla-phy}), and 
their apparent sizes measured with a $\sim 1\arcsec$ resolution 
are comparable to the X-ray knots (Table~\ref{tab-feknots}). 
So, if we naively accept these numbers, the NIR Fe-rich knots are expected to be optically thick for $\gamma$ rays generated from the radioactive decay from $^{56}$Ni to $^{56}$Fe. 
However, the structure of the NIR Fe-rich knots are not resolved. 
It is possible that they consist of multiple, compact subknots with 
relatively low column densities, 
enabling the $\gamma$ rays to escape primarily from the knot.
The {\it HST} images in Figure~\ref{fig-counter} 
show that indeed the intrinsic sizes of the NIR knots could be much smaller 
than their apparent sizes in the deep \feii\ image.
Furthermore, the ejecta clumps are destroyed 
by the passage of the reverse shock \citep[e.g., see Figure 4 of ][]{slavin20}, 
so the initial sizes of the NIR and X-ray knots could have been considerably smaller.  
Hence, the Ni bubble effect may not be significant for 
either type of Fe-rich knots. 
We will conduct subsequent studies in a forthcoming paper. 

Before leaving this section, 
it is worth noticing that our observation has been done 
towards bright knots in the deep \feii\ image 
obtained by using a narrow band filter with a bandwidth of $\pm 2800~\kms$
(see Section~\ref{sec-obs}),
so there is a possibility that there are additional  
Fe-rich knots outside this radial velocity range.
However, previous studies have shown that the diffuse Fe ejecta typically 
have radial velocities in the range of  
$v_{\rm r}=-3000$ to $-1000~\kms$
\citep{del10,tsuchioka22}. 
Therefore, it is likely that the majority of Fe-rich ejecta knots are included in the 
\feii\ image of Figure \ref{fig-se_detail}, and 
our result provides insight into the distribution of the Fe-rich ejecta knots. 

\subsection{Knots with only \hei\ 1.083~\micron\ lines} \label{sec-dis-he}

\begin{figure*}
\center
\includegraphics[width=0.9\textwidth]{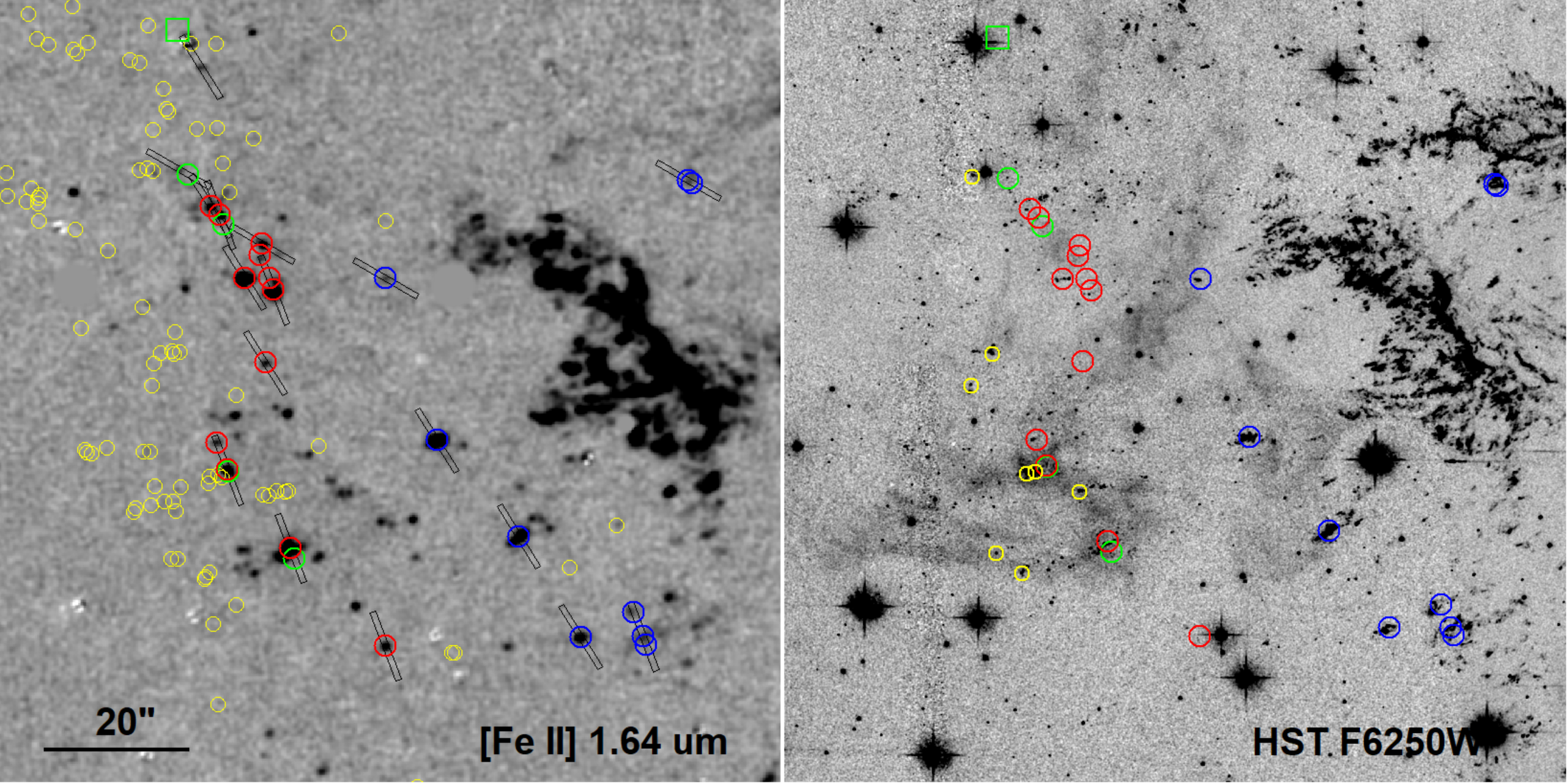}
\caption{
Distribution of optical NKs (yellow circles) in the catalog 
of \cite{ham08} shown on the deep \feii\ and \hst\ F625W images. 
North is up, and east is to the left.
The circles/squares and their colors are the same as in Figure \ref{fig-se_detail}.
The black bars in the deep \feii\ image represent the locations of the 10\arcsec-long MOS slits.
In the F6250W image, only the bright NK knots 
with an F6250W flux larger than $6\times 10^{-17}$ erg cm$^{2}$ s$^{-1}$ are shown. 
Note that the positions of the slit and knot symbols are shifted to the epochs
of the background images, i.e., 2013 Sept. for the deep \feii\ image  
and 2004 Mar. for the {\it HST} image. 
} \label{fig-heknots}
\end{figure*}

We detected seven knots that show only \hei\ 1.083~\micron\ lines.
The \hei\ 1.083 \micron\ line intensities of these \hei\ knots are not particularly strong 
in comparison with other metal-rich ejecta knots (Figure \ref{fig-fxvsfx}).
But, since \feii\ 1.257 \micron\ line is not detected, 
they are noticeable by their very high \heioverfeii, 
which is 1--2 orders of magnitude higher than the typical S-rich and Fe-rich ejecta 
knots (see the bottom frame of Figure \ref{fig-class}).
Table \ref{tab-heknots} lists the knots with \heioverfeii\ greater 
than 8 where we can see that the majority are \hei\ knots.

\hei\ knots are detected along the forward shock front, 
from the southern base of the NE jet to the Fe K plume area (see Figure \ref{fig-pvd}).
The radial velocities of \hei\ knots are spread over a very wide range,
from $-3300~\kms$ to $+3200~\kms$ (Figure \ref{fig-histogram} and Table \ref{tab-heknots}).
Since most of them have very high radial velocities, 
they are probably not circumstellar material.
Their spectral properties are also different from the 
low-velocity circumstellar clumps in Cas A, 
which are called {\em quasi-stationary flocculi} (QSFs).
QSFs are He and N-enriched circumstellar
material ejected by the progenitor before SN explosion 
and have been shocked by the SN blast wave recently 
(\citealt{pei71a,mckee75}; see also \citealt{koo23} and references therein). 
In NIR, QSFs show strong \hei\ lines together with  
moderately strong \feii\ and \sii\ lines and also weak  
Pa$\gamma$ and \ci\ 0.985~\micron\ lines \citep{chevalier78,ger01,lee17}.  
The ratio of Pa$\gamma$ to \hei\ 1.083~\micron\ is typically $\sim 0.03$ 
(see \citealt{koo23}).
For comparison, all \hei\ knots  
show strong ($>10^{-15}$ erg cm$^{-2}$ s$^{-1}$) 
\hei\ lines without \feii\ or \sii\ lines, and about half of them also show 
weak \ci\ 0.985~\micron\ line. 

What's the origin of \hei\ knots? The first possibility that we can consider  
is that they are debris of the He-rich envelope material of the progenitor star 
expelled by the SN explosion, like high-velocity 
N-rich knots (NKs) detected in optical studies 
\citep{fes87,fes88,fes91,fes01,ham08, fes16}.
The spectra of NKs are dominated by strong \nii\ $\lambda\lambda$6548, 6583 lines 
in the wavelength region 4500--7200 \AA.
For most NKs, H$\alpha$ line is either much weaker than \nii\ lines or
undetected, so NKs are considered as important evidence 
that the Cas A progenitor had a thin N-rich and H-poor envelope at the time of SN explosion. 
A deep, ground-based \nii\ imaging observations detected 50 NKs, 
whereas {\it HST} observations revealed numerous NKs distributed around the remnant.
In particular, in the eastern area where \hei\ knots have been detected, 
many NKs had been detected (Figure \ref{fig-heknots}; 
see also Figure 5 of \citealt{fes01} or Figure 9E of \citealt{fes16}). 
The radial velocities of the bright NK knots in the eastern area are in the 
range of $-2000$ to +2500~\kms~\citep{fes01}.
%+2450, +1400, 0 to -2000, +230, +40, +400 (Fesen 2001).
%4 He knots + radia l velocity similar
%3 FMKs -3000 km/s 
So the association of \hei\ knots with NKs seems plausible considering that 
He abundance might be enhanced 
in the CNO-processed, N-rich envelope of the progenitor star. 
Apparently, however, none of the \hei\ knots 
has an optical NK counterpart (Table \ref{tab-heknots}; see \S~\ref{sec-res-opt}). 
Another difficulty in connecting \hei\ knots to NKs
is the non-detection of \hei\ $\lambda$5876 line in NKs \citep{fes91,fes01}. 
The recombination emissivity of \hei\ $\lambda$5876 line 
is a factor of 2--10 smaller than that of \hei\ 1.083 \micron\ line 
depending on electron density 
($10^2$--$10^4$ cm$^{-3}$) in the Case B approximation \citep[e.g.,][]{dra03}.
The flux will be further reduced 
by interstellar extinction. Hence, \hei\ $\lambda$5876 line is expected to 
be much weaker than \hei\ 1.083 \micron\ line, but it is not obvious 
if that can explain the non-detection of \hei\ $\lambda$5876 lines.
An interesting possibility is that the 
\hei\ knots originated from the lower portion 
of the He-rich envelope of the progenitor star where N is depleted. 
If a large amount of carbon is dredge-up by convection, 
nitrogen could be depleted because $^{14}$N mixed into the lower layer by convection participates the
$^{14}\mathrm{N}(\alpha,\gamma)^{18}\mathrm{F}(\beta^+)^{18}\mathrm{O}(\alpha,
\gamma)^{22}\mathrm{Ne}$ reaction. 
Such a $^{14}$N-depleted region formed by convective
dredge-up is often found at the pre-SN
stage in massive star models (Y. Jung et al, in preparation). 
Observationally, our slits are positioned toward \feii\ line-emitting knots by
design. So it is possible that we are picking up He-rich, N-depleted knots that are located
closer to the explosion center than the optical NKs, but which coincidentally overlap with
the same region of sky as the Fe-emitting knots (Figure \ref{fig-heknots}).
Future studies may further explore the possibility.

\begin{figure*}
\begin{center}
\includegraphics[width=0.9\textwidth]{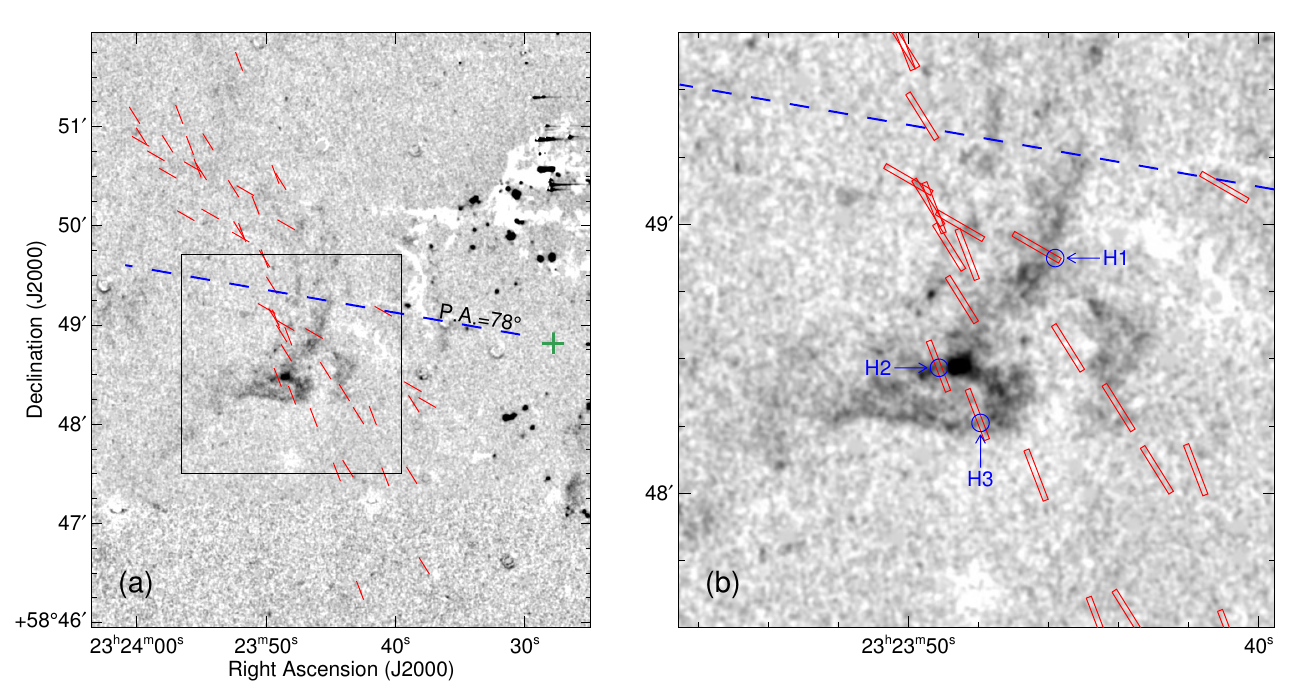}
\caption{
	(a) Continuum-subtracted \ha\ image of the eastern area of Cas A.
	The symbols and lines are the same as in Figure \ref{fig-slitpos}.
      (b) An enlarged view of the boxed area in (a).
      	The blue circles represent	the central positions of the H emission features:
	H1, H2, and H3 in Table~\ref{tab-hi}.
	The image has been produced from the Isaac Newton Telescope (INT)
	Wide Field Survey (WFS) data obtained in 2005 October.
} \label{fig-ha}
\end{center}
\end{figure*}

Another possibility is that they are metal-enriched SN ejecta
with relatively high He abundance. 
We note that some S-rich and Fe-rich knots
have very high \heioverfeii, as high as the \hei\ knots,
and that the majority of \hei\ knots have 
\sii\ 1.03 \micron\ and \ci\ 0.985~\micron\ flux upper limits that are not particularly 
strong (Table \ref{tab-heknots}).
This explanation
also appears to be consistent with that the half of He knots seem to be FMK-like knots
in optical classifications (Figure~\ref{fig-opt}).
The overabundance of low atomic elements
could be attributed to significant mixing among the nucleosynthetic layers
as pointed out in previous X-ray and optical studies \citep{fes91,fes96}.
From optical spectroscopy,
\citet{fes91} reported the detection of a hybrid ejecta knot
showing O and S emission lines along with \nii\ and \ha\ emission lines in the NE jet region,
and they called it a `mixed emission knot' (MEK).
Additional optical studies towards the outer region of the remnant
have further detected several MEKs,
which suggested  chemical mixing among the nucleosynthetic layers
between the explosive O-burning layer and the N-rich photospheric envelop of the progenitor
\citep{fes96,fes01}.
Yet another possibility is that \hei\ knots are the remains of 
the complete Si burning with $\alpha$-rich freezeout in 
the innermost region of the progenitor as helium in Fe-rich knots \citep[see also][]{lee17}. 
But it seems to be difficult for this scenario to explain 
the difference in radial velocities between the \hei\ knots  
and the Fe-rich knots.

It is also worth noticing that Figure~\ref{fig-pvd} does not represent 
the overall distribution of \hei\ knots.
Since our observation was done toward ejecta knots bright in 
\feii\ emission, the detection of \hei\ knots was unexpected. 
\hei\ knots were detected because they fell on the slit by chance. 
In order to see the overall distribution \hei\ knots,
which is essential for understanding their origin,
we need \hei\ 1.083~\micron\ narrow-band observations of the Cas A remnant.

\subsection{Extended H emission Features Associated with East Cloud} \label{sec-dis-ism}

In addition to the 67 SN ejecta knots, we also detected
three ``extended H emission features'' in the outer eastern region
(Section~\ref{sec-iden}),
and their physical properties are listed in Table~\ref{tab-hi}.
They show only unresolved (FWHM $\simlt 9$ \AA) \hi\ lines,
e.g., \pg\ 1.094~\micron\ and \pb\ 1.282~\micron\ lines,
without any emission lines from metallic elements,
and have radial velocities between $-20$ \kms\ and $-30~\kms$,
indicating that they are of CSM/ISM origin.
We can estimate the line-of-sight extinction to the H-emitting gas
using the flux ratio of \hi\ emission lines.
All three features show Pa$\beta$ line, and
the brightest one (H2) also shows Pa$\gamma$ line.
The observed \hi\ line ratio, $F$(Pa$\gamma$)/$F$(Pa$\beta$),
is $0.41\pm0.09$.
By comparing with the unreddened line ratio of 0.56
at $T_{\rm e} = 10^4$~K and $n_{\rm e} = 10^4$~cm$^{-3}$
\citep[][]{hum87},
we obtained $N_{\rm H}$ of $7.6^{+6.4}_{-5.1} \times 10^{21}$ cm$^{-2}$
which is corresponding to $A_{\rm V}$ of $4.1^{+3.4}_{-2.7}$ mag.
The extinction-corrected  $F$(\feii\ 1.257)/$F$(\pb) is $<0.2$
($2\sigma$ upper limit),
which is at least an order of magnitude smaller than
that in the typical shock-excited CSM/ISM
detected in the several Galactic SNRs
\citep[2--20;][and references therein]{koo15}.
This low $F$(\feii\ 1.257)/$F$(\pb) indicates that
the \hi\ lines are associated with a photoionized gas.

We noticed that the positions of the extended H emission features are coincident with
the region where a bright, diffuse optical cloud is detected
(Figure~\ref{fig-ha}).
The diffuse cloud in the eastern region
(hereafter ``East Cloud''), located 2\farcm7 east from the center of the remnant, 
was first identified in an early optical photograph taken by R. Minkowski
\citep{min68,van71},
and it has a triangular shape with a size of $\sim 1\arcmin$
(see also the right panel of Figure~\ref{fig-ha}).
In a deep \ha\ narrow-band image, however,
it turned out to be more extended and diffuse 
(see Figure~10 in \citealt{fes01} or Figure~2 in \citealt{weil20}).
The follow-up optical spectroscopy showed that
the East Cloud emits strong \ha\ emission line accompanied by
relatively weak \oiii\ $\lambda\lambda$ 4959, 5007, \nn\ $\lambda\lambda$ 6548, 6583,
and \sii\ $\lambda\lambda$ 6716, 6725 lines \citep{fes87,weil20}.

The origin of the East Cloud as well as
its physical association with Cas A has been controversial.
The early optical studies suggested that
it is a small, diffuse \hii\ region adjacent to the remnant \citep{min68,van71}.
However, the lack of OB stars around the eastern cloud suggested that
it is a diffuse CSM/ISM nearby Cas A excited by the UV/X-ray emission
from the SN outburst \citep{pei71b,van71,weil20}.
In this scenario,
the East Cloud could be either the diffuse ISM
surrounding the progenitor star \citep{min68,van71,pei71a,pei71b}
or the diffuse CSM blown out from the progenitor star
during its red supergiant phase \citep{fes87,che03,weil20}.
The peculiar structure of the East Cloud,
its projected proximity to the NE jet, and distinct 
spectroscopic properties relative to the surrounding diffuse emissions
are indirect evidence suggesting the physical association
of the East Cloud with the remnant \citep{fes87,fes96,fes06,weil20}, while 
the presence of brightened ejecta knots matching the East Cloud’s emission structure
provides strong evidence that the East Cloud is physically adjacent to Cas A \citep{weil20}.

According to our spectroscopy, 
the mean $v_{\rm LSR}$ of the \pb\ lines weighted by 1/$\sigma_{v}^{2}$
(where $\sigma_{v}$ is the $1\sigma$ uncertainty of the central velocity)
is $-22\pm5~\kms$ (see Table~\ref{tab-hi}).
Considering the systematic uncertainty from the wavelength calibration
($\sim 12~\kms$; see Section~\ref{sec-obs}),
the $v_{\rm LSR}$ of the East Cloud is $-22\pm13~\kms$.
This is consistent with the previous result:
$v_{\rm LSR}=-20\pm75~\kms$ from a few bright optical lines \citep{fes87}
and $v_{\rm LSR}=-13\pm10~\kms$ from \ha\ lines \citep{ala14}.
Previous radio observations, however, showed that
Cas A is located at the far-side of the Perseus spiral arm implying 
$v_{\rm LSR}\sim -50~\kms$ for Cas A \citep[][and references therein]{zho18}.
Furthermore, recent NIR observations for the southern optical QSF (``Knot 24'')
reported the detection of narrow (${\rm FWHM} \sim 8~\kms$) \feii\ lines
from unshocked pristine CSM with $v_{\rm LSR}\simeq-50~\kms$
\citep{koo20},
which is consistent with the systematic velocity center $v_{\rm LSR}\simeq -50~\kms$ of Cas A.
So the velocity of the East Cloud ($v_{\rm LSR}\sim -22~\kms$) is 
considerably different from the systemic velocity of the Cas A SNR.
This might indicate that the East Cloud is not physically associated with the remnant,
but is located at $\sim 2$~kpc from us (more than 1~kpc closer than the remnant)
assuming the flat Galactic rotation model
with IAU standard ($R_{0}=8.5$~kpc and $v_{0}=220~\kms$).
Relatively low line-of-sight extinction of the East Cloud
($A_{\rm V} \sim 4$~mag) derived from their \hi\ lines
compared to that toward the remnant
($A_{\rm V}=6$--$10$~mag; see Figure~\ref{fig-ext})
seems to support this possibility.
On the other hand, if the East Cloud is the CSM material ejected 
from the progenitor star as suggested by \citet{che03} and \citet{weil20},
the velocity difference ($\sim 30~\kms$) should represent 
the line-of-sight component of the ejection velocity of the CSM.
And, considering that
the East Cloud is located at the eastern outer edge of the remnant,
the ejection velocity should have been much higher than $30$~\kms,   
or than the typical wind velocity of red supergiant stars, e.g., $10$--$20~\kms$
\citep{smith14}.

\section{Summary} \label{sec-sum}

The Cas A SNR has a quite complex structure, manifesting 
the violent and asymmetric explosion. 
Two representative features are the NE jet and the Fe K plume 
in the outer eastern area of the SNR.
The NE jet is a stream of ejecta material dominated by 
intermediate mass elements beyond the SN blast wave, 
and the Fe K plume is a plume of X-ray emitting 
hot gas dominated by Fe ejecta outside the main ejecta shell. 
These two features are expanding much faster than the main ejecta shell 
of the SNR, suggesting turbulent convection and hydrodynamic 
instabilities in the early stages of the SN explosion.
We carried out NIR (0.95--1.75~\micron) MOS spectroscopy of the 
NE jet and Fe K plume regions of the Cas A SNR using MMIRS. 
In the two-dimensional spectra of 52 MOS slits, which were positioned on the bright knots  
in the deep \feii\ 1.64 $\mu$m image of \cite{koo18}, 
a total of 67 knots have been identified.
All knots show at least one of the following strong lines: 
\siii\ 0.983 $\mu$m, \hei\ 1.083 $\mu$m, \sii\ 1.03 $\mu$m, and \feii\ 1.257 $\mu$m lines. 
And about one third of the knots also show strong \pii\ 1.189 $\mu$m line. 
We find that the knots in different areas show distinctively different ratios of these lines, 
suggesting their different elemental composition. 
The NIR lines are emitted from shocked gas, so  
their intensities depend on shock environments as well as the elemental composition of the knots, 
e.g., density and velocity of the knots, density contrast between the knot and the ambient gas, 
ambient pressure. In metal rich ejecta knots, the elemental composition also 
profoundly affects the cooling function and therefore the physical structure of the shocked gas 
 \citep[e.g.,][]{raymond19}.
Hence, it would be a formidable task to derive the elemental composition 
from just the NIR spectra.
In this work, we simply classify the knots into three groups 
based on the relative strengths of \sii, \feii, and \hei\ lines,  i.e., 
S-rich, Fe-rich, and \hei\ knots, and explore 
the origins of these knots and their connection to the explosion 
dynamics of the Cassiopeia A supernova. 
We summarize our main results as follows.

\begin{enumerate}
  
\item
The NE jet is dominated by S-rich knots. 
There are no Fe-rich knots without \sii\ lines. 
The knots show weak or no \pii\ lines, which clearly
differentiates them from the S-rich ejecta in the main ejecta shell 
that have strong \pii\ lines. 
The low P abundance inferred from their low \pii/\feii\ line ratios
indicate that these S-rich knots were produced below the explosive Ne burning layer,
which is consistent with the results of previous studies.
Our results do not support the jet-induced explosion model, 
but it is also not clear if the NE jet can be explained by 
the neutrino-driven explosion model.

\item
In the Fe K plume area, 
along the forward shock front just outside 
the boundary of the diffuse X-ray emitting Fe ejecta,
we detected Fe-rich knots showing only \feii\ lines with or 
without \hei\ line.
These NIR Fe-rich ejecta knots are expanding with velocities considerably 
higher than the X-ray Fe ejecta.
The spatial and kinematic relations support the 
physical association of these dense NIR Fe-rich knots with the X-ray diffuse 
Fe ejecta produced by explosive complete Si burning with $\alpha$ freezeout.
We suggest that the initial density distribution of the Fe ejecta 
produced in the innermost region was very inhomogeneous and that 
the dense knots were ejected with the diffuse Fe ejecta but   
decoupled after crossing the reverse shock.

\item
We also detected several \hei\ knots emitting only \hei\ 1.083~\micron\ lines 
with or without very weak \ci\ 0.985 $\mu$m lines.
They are detected along the forward shock front 
from the southern base of the NE jet to the Fe K plume area.
The origin of these He-rich knots is unclear. 
They are likely the debris of the He-rich layer above the carbon-oxygen core of 
the progenitor expelled during the SN explosion, but they could be 
also metal-enriched SN ejecta with relatively high He abundance  
or the remains of the explosive complete Si burning in $\alpha$-rich freezeout.
The detection of \hei\ knots was 
unexpected because our MMT observations were directed toward  
ejecta knots bright in \feii\ emission. So 
an imaging observation is needed to reveal the distribution of \hei\ knots.

\item
In addition to the 67 SN ejecta knots,
we also detected three extended H emission features 
associated with the diffuse cloud, the East Cloud, in the eastern area well beyond the SNR boundary. 
They show only narrow H recombination lines with low line of sight velocities ($\sim -20$~\kms), indicating  
that the lines are arising from photoionized CSM/ISM.
Their velocities are substantially different from the systematic velocity of the Cas A ($v_{\rm LSR}\sim -50$~\kms), 
and it implies that, if the East Cloud is a CSM ejected from the progenitor star of the Cas A SN, 
the ejection velocity should have been much higher than 30~\kms\ considering its location.

\end{enumerate}

\begin{acknowledgments}
%\acknowledgements
This work was supported by K-GMT Science Program (PID: MMT-2017B-4) of
Korea Astronomy and Space Science Institute (KASI).
BCK was supported by Basic Science Research Program through the National Research Foundation of
Korea(NRF) funded by the Ministry of Science, ICT and future Planning 
(2020R1A2B5B01001994).
Observations reported here were obtained at the MMT Observatory,
a joint facility of the Smithsonian Institution and the University of Arizona.
\end{acknowledgments}

%\facility{facility ID}
\facilities{MMT, UKIRT} 
\software{IDL, Python}

\bibliographystyle{yahapj}
\bibliography{references}

%\appendix
%\section{appendix section}

%===== Tables =====
%\clearpage
\begin{deluxetable}{lccccc}
%\vspace{3.0truecm}
\tabletypesize{\scriptsize}
\tablewidth{0pt}
\tablecolumns{6}
\tablecaption{Log of MOS Spectroscopy \label{tab-log}}
\tablehead{
\colhead{Mask}	&   \colhead{Number of Slits}	&	\colhead{Grism}	&
\colhead{Filter}	&   \colhead{P.A.}	&
\colhead{Exposure Time}	\\
	&   	& 	&   &	\colhead{(degree)}	& \colhead{(sec)}	
}
\decimalcolnumbers
\startdata
MASK 1	&   20  &	J		&	zJ	&	32	&	$300 \times 8$	\\
		&       &	H3000	&	H	&	32	&	$300 \times 4$	\\[5pt]
MASK 2	&   17  &	J		&	zJ	&	21	&	$300 \times 12$	\\[5pt]
MASK 3	&   15  &	J		&	zJ	&	60	&	$300 \times 12$	\\
\enddata
\tablecomments{
(1) mask name; (2) number of MOS slits in each mask; 
(3) grism; (4) filter; (5) position angle of slits 
measured counterclockwise from north to east on the plane of the sky; 
(6) exposure time per frame (sec) $\times$ number of dithering. 
}
\end{deluxetable}

\setlength{\belowdeluxetableskip}{-0.8truecm}
\begin{deluxetable}{lccc|lccc}
%\vspace{-0.3truecm}
\tabletypesize{\scriptsize}
\tablewidth{0pt}
\tablecolumns{8}
\tablecaption{Slit Positions\label{tab-slit}}
\tablehead{
\colhead{Mask}	&   \colhead{Slit No.}	&
\colhead{$\alpha(J2000)$}	&   \colhead{$\delta(J2000)$} & 
\colhead{\vl\, \ \ \ Mask}	&   \colhead{Slit No.}	&
\colhead{$\alpha(J2000)$}	&   \colhead{$\delta(J2000)$}	  	
}
%\decimalcolnumbers
\startdata
MASK 1 & 05  &  23:24:00.342  &  +58:51:06.45  &  MASK 3 & 07  &  23:23:59.854  &  +58:50:50.70  \\
       &  06  &  23:23:59.797  &  +58:50:53.89  &       &  08  &  23:23:58.676  &  +58:50:42.03  \\
       &  07  &  23:23:57.769  &  +58:50:49.28  &       &  09  &  23:23:57.739  &  +58:50:31.51  \\
       &  08  &  23:23:54.621  &  +58:50:50.54  &       &  10  &  23:23:55.808  &  +58:50:35.53  \\
       &  09  &  23:23:55.148  &  +58:50:32.56  &       &  11  &  23:23:56.335  &  +58:50:05.89  \\
       &  10  &  23:23:52.626  &  +58:50:22.26  &       &  12  &  23:23:54.448  &  +58:50:07.00  \\
       &  11  &  23:23:48.962  &  +58:50:26.84  &       &  13  &  23:23:51.720  &  +58:50:21.20  \\
       &  12  &  23:23:52.200  &  +58:49:57.42  &       &  14  &  23:23:52.053  &  +58:49:53.03  \\
       &  13  &  23:23:50.194  &  +58:49:40.02  &       &  15  &  23:23:48.541  &  +58:50:00.63  \\
       &  14  &  23:23:49.642  &  +58:49:24.08  &       &  16  &  23:23:50.035  &  +58:49:09.96  \\\
       &  15  &  23:23:49.462  &  +58:49:04.85  &       &  17  &  23:23:48.548  &  +58:48:59.68  \\
       &  16  &  23:23:48.859  &  +58:48:54.93  &       &  18  &  23:23:46.346  &  +58:48:54.93  \\
       &  17  &  23:23:48.487  &  +58:48:43.22  &       &  19  &  23:23:40.964  &  +58:49:08.44  \\
       &  18  &  23:23:45.436  &  +58:48:32.51  &       &  22  &  23:23:38.661  &  +58:48:22.95  \\
       &  19  &  23:23:43.979  &  +58:48:19.23  &       &  23  &  23:23:37.549  &  +58:48:13.42  \\
       &  20  &  23:23:42.877  &  +58:48:05.38  &  && \\
       &  21  &  23:23:38.578  &  +58:48:12.57  &   && \\
       &  22  &  23:23:43.680  &  +58:47:33.33  &   && \\
       &  24  &  23:23:38.731  &  +58:47:29.37  &   && \\
       &  28  &  23:23:37.742  &  +58:46:34.72  &   && \\
MASK 2 & 03  &  23:23:52.229  &  +58:51:39.22  &&&\\
       &  04  &  23:23:56.865  &  +58:51:07.09  &&&\\
       &  05  &  23:23:56.026  &  +58:50:48.34  &&&\\
       &  06  &  23:23:55.432  &  +58:50:34.78  &&&\\
       &  07  &  23:23:49.371  &  +58:50:31.02  &&&\\
       &  08  &  23:23:50.915  &  +58:50:12.42  &&&\\
       &  09  &  23:23:51.953  &  +58:49:55.88  &&&\\
       &  10  &  23:23:50.194  &  +58:49:40.02  &&&\\
       &  12  &  23:23:49.296  &  +58:49:03.66  &&&\\
       &  13  &  23:23:48.340  &  +58:48:53.28  &&&\\
       &  14  &  23:23:49.155  &  +58:48:28.37  &&&\\
       &  15  &  23:23:48.034  &  +58:48:17.59  &&&\\
       &  16  &  23:23:46.344  &  +58:48:04.12  &&&\\
       &  17  &  23:23:41.761  &  +58:48:05.40  &&&\\
       &  18  &  23:23:44.543  &  +58:47:31.34  &&&\\
       &  19  &  23:23:40.775  &  +58:47:28.46  &&&\\
       &  23  &  23:23:42.721  &  +58:46:19.80  &&&\\
\enddata
\end{deluxetable}

\setlength{\belowdeluxetableskip}{0pt}

%\clearpage
%\startlongtable
\begin{deluxetable}{ccccrrrrcrc}
\tabletypesize{\tiny}
\tablewidth{0pt}
\tablecolumns{10}
\tablecaption{Physical Parameters of 67 Ejecta Knots \label{tab-knots}}
\tablehead{
\colhead{Knot No.}  &  \colhead{ID}  & \colhead{$\alpha$(J2000)} & \colhead{$\delta$(J2000)}   &
\colhead{P.A.}	&   \colhead{$d_{\rm rad}$}	&
\colhead{FWHM}	&   \colhead{$v_{\rm rad}$} &
\colhead{$N_{\rm H}$}   &   \colhead{Knot}  & K2018 No.\tablenotemark{a} \\
\colhead{}	& \colhead{} &   	& 	&   \colhead{($\degr$)}  &  \colhead{($\arcmin$)}  &
\colhead{(\kms)}	&   \colhead{(\kms)}    &
\colhead{($10^{22}$ cm$^{-2}$)} &   \colhead{Type} & 
}
\decimalcolnumbers
\startdata
 1	&	2-07-2	&	23:23:49.49 & +58:50:33.4 &  58.3	&	3.30	&	 277 ( 8) & $+2336~( 2)$	&	1.42	&	S	&	169B \\ 
 2	&	2-04-1	&	23:23:56.87 & +58:51:07.1 &  58.6	&	4.41	&	 292 (34) & $ +420~(15)$	&	1.85	&	S	&	182 \\ 
 3	&	2-07-1	&	23:23:49.37 & +58:50:31.0 &  58.7	&	3.27	&	 298 ( 2) & $ +141~( 1)$	&	1.42	&	S	&	169A \\ 
 4	&	1-11-1	&	23:23:48.96 & +58:50:26.8 &  59.3	&	3.19	&	 318 ( 6) & $ +272~( 3)$	&	1.41	&	S	&	168 \\ 
 5	&	1-08-1	&	23:23:54.62 & +58:50:50.5 &  59.8	&	4.02	&	 249 ( 1) & $ +124~( 1)$	&	1.60	&	S	&	180 \\ 
 6	&	1-05-2	&	23:24:00.52 & +58:51:08.7 &  61.2	&	4.83	&	 247 (26) & $ +172~(11)$	&	1.75	&	S\tablenotemark{b}	&	$\cdots$ \\ 
 7	&	2-05-1	&	23:23:56.03 & +58:50:48.3 &  61.5	&	4.16	&	 214 (16) & $ +428~( 7)$	&	1.74	&	S	&	181 \\ 
 8	&	1-05-1	&	23:24:00.34 & +58:51:06.5 &  61.5	&	4.79	&	 264 (19) & $ +748~( 9)$	&	1.72	&	S	&	184 \\ 
 9	&	1-07-1	&	23:23:57.77 & +58:50:49.3 &  62.7	&	4.37	&	 297 (17) & $ +832~(10)$	&	1.88	&	S	&	200 \\ 
10	&	1-06-2	&	23:24:00.19 & +58:50:58.8 &  62.7	&	4.72	&	 174 (11) & $ +452~( 4)$	&	1.77	&	S	&	$\cdots$ \\ 
11	&	1-06-1	&	23:23:59.80 & +58:50:53.9 &  63.3	&	4.63	&	 286 ( 3) & $  +10~( 2)$	&	1.73	&	S	&	207 \\ 
12	&	3-13-1	&	23:23:51.72 & +58:50:21.2 &  63.7	&	3.46	&	 248 ( 5) & $ +429~( 3)$	&	1.55	&	S	&	188 \\ 
13	&	2-06-1	&	23:23:55.43 & +58:50:34.8 &  63.8	&	3.99	&	 339 ( 7) & $ +541~( 4)$	&	1.58	&	S	&	195A \\ 
14	&	3-10-1	&	23:23:55.81 & +58:50:35.5 &  64.0	&	4.04	&	 369 ( 9) & $ +519~( 4)$	&	1.61	&	S	&	195B \\ 
15	&	3-07-1	&	23:23:59.85 & +58:50:50.7 &  64.0	&	4.62	&	 311 ( 7) & $ +110~( 4)$	&	1.73	&	S	&	206 \\ 
16	&	1-09-1	&	23:23:55.15 & +58:50:32.6 &  64.1	&	3.94	&	 227 ( 8) & $ +483~( 3)$	&	1.58	&	S	&	194 \\ 
17	&	1-10-1	&	23:23:52.62 & +58:50:22.3 &  64.3	&	3.57	&	 283 ( 3) & $ +490~( 2)$	&	1.66	&	S	&	189 \\ 
18	&	3-08-1	&	23:23:58.68 & +58:50:42.0 &  64.8	&	4.42	&	 245 (28) & $ +625~(10)$	&	1.73	&	S	&	201 \\ 
19	&	2-08-1	&	23:23:50.92 & +58:50:12.4 &  65.2	&	3.30	&	 421 (61) & $ +799~(26)$	&	1.49	&	S	&	187A \\ 
20	&	2-08-2	&	23:23:50.90 & +58:50:12.0 &  65.2	&	3.29	&	 271 (39) & $ -230~(17)$	&	1.49	&	S	&	187B \\ 
21	&	3-15-2	&	23:23:49.10 & +58:50:03.1 &  66.0	&	3.02	&	 312 (28) & $ +329~(14)$	&	1.46	&	S\tablenotemark{b}	&	$\cdots$ \\ 
22	&	3-15-1	&	23:23:48.54 & +58:50:00.6 &  66.1	&	2.94	&	 441 (16) & $+2485~( 8)$	&	1.46	&	S	&	$\cdots$ \\ 
23	&	3-09-1	&	23:23:57.74 & +58:50:31.5 &  66.3	&	4.24	&	 227 (10) & $ -234~( 4)$	&	1.66	&	Fe	&	197 \\ 
24	&	3-12-1	&	23:23:54.45 & +58:50:07.0 &  69.4	&	3.69	&	 302 ( 6) & $ -696~( 3)$	&	1.64	&	S	&	220 \\ 
25	&	1-12-1	&	23:23:52.20 & +58:49:57.4 &  70.2	&	3.36	&	 252 ( 3) & $ -600~( 2)$	&	1.65	&	S	&	214A \\ 
26	&	1-12-3	&	23:23:52.16 & +58:49:56.9 &  70.3	&	3.35	&	 241 (15) & $+3242~( 8)$	&	1.65	&	He	&	$\cdots$ \\ 
27	&	2-09-1	&	23:23:51.95 & +58:49:55.9 &  70.5	&	3.32	&	 317 ( 6) & $ -515~( 2)$	&	1.65	&	S	&	214B \\ 
28	&	3-11-1	&	23:23:56.34 & +58:50:05.9 &  70.9	&	3.91	&	 304 (11) & $-1199~( 5)$	&	1.62	&	S	&	224 \\ 
29	&	1-12-2	&	23:23:51.87 & +58:49:53.3 &  71.1	&	3.30	&	 256 (11) & $-2116~( 6)$	&	1.65	&	S\tablenotemark{b}	&	$\cdots$ \\ 
30	&	3-11-2	&	23:23:55.84 & +58:50:03.7 &  71.1	&	3.84	&	 247 ( 9) & $ +752~( 4)$	&	1.62	&	S	&	222 \\ 
31	&	3-14-1	&	23:23:52.05 & +58:49:53.0 &  71.3	&	3.32	&	 281 ( 3) & $-2125~( 2)$	&	1.65	&	S	&	212 \\ 
32	&	3-14-2	&	23:23:51.78 & +58:49:51.8 &  71.4	&	3.28	&	 217 ( 8) & $-3170~( 4)$	&	1.62	&	S\tablenotemark{b}	&	$\cdots$ \\ 
33	&	2-09-2	&	23:23:51.72 & +58:49:51.2 &  71.6	&	3.27	&	 249 ( 9) & $-3152~( 3)$	&	1.62	&	He	&	$\cdots$ \\ 
34	&	1-13-1	&	23:23:50.19 & +58:49:40.0 &  73.8	&	3.02	&	 250 ( 8) & $-2017~( 4)$	&	1.49	&	S	&	226 \\ 
35	&	1-14-2	&	23:23:50.05 & +58:49:29.2 &  77.0	&	2.96	&	 187 (10) & $+2314~( 6)$	&	1.48	&	He	&	$\cdots$ \\ 
36	&	3-19-1	&	23:23:40.96 & +58:49:08.4 &  79.5	&	1.74	&	 358 (35) & $-1954~(14)$	&	1.57	&	S	&	229 \\ 
37	&	3-19-2	&	23:23:40.90 & +58:49:08.1 &  79.6	&	1.73	&	 286 (22) & $-2411~( 9)$	&	1.57	&	S	&	$\cdots$ \\ 
38	&	3-16-2	&	23:23:49.86 & +58:49:09.2 &  83.4	&	2.88	&	 211 (14) & $-3312~( 6)$	&	1.47	&	He	&	$\cdots$ \\ 
39	&	1-15-1	&	23:23:49.46 & +58:49:04.9 &  84.7	&	2.82	&	 247 (13) & $-1286~( 5)$	&	1.42	&	Fe	&	235 \\ 
40	&	2-12-1	&	23:23:49.30 & +58:49:03.7 &  85.1	&	2.80	&	 268 (13) & $-1282~( 5)$	&	1.42	&	Fe	&	234 \\ 
41	&	2-12-2	&	23:23:49.23 & +58:49:02.4 &  85.5	&	2.79	&	 271 (23) & $ -269~(11)$	&	1.42	&	He	&	$\cdots$ \\ 
42	&	3-17-1	&	23:23:48.55 & +58:48:59.7 &  86.3	&	2.70	&	 210 (17) & $-1616~( 8)$	&	1.39	&	Fe	&	232 \\ 
43	&	2-13-3	&	23:23:48.58 & +58:48:58.1 &  86.9	&	2.70	&	 259 (17) & $-1987~( 5)$	&	1.39	&	Fe	&	$\cdots$ \\ 
44	&	3-18-1	&	23:23:46.35 & +58:48:54.9 &  87.8	&	2.41	&	 261 (14) & $-1319~( 6)$	&	1.47	&	S	&	238 \\ 
45	&	2-13-2	&	23:23:48.42 & +58:48:55.0 &  88.0	&	2.68	&	 235 ( 8) & $-1165~( 3)$	&	1.37	&	Fe	&	$\cdots$ \\ 
46	&	1-16-1	&	23:23:48.86 & +58:48:54.9 &  88.0	&	2.73	&	 242 ( 3) & $-1189~( 2)$	&	1.39	&	Fe	&	241 \\ 
47	&	2-13-1	&	23:23:48.34 & +58:48:53.3 &  88.6	&	2.66	&	 328 ( 4) & $-1159~( 2)$	&	1.37	&	Fe	&	239 \\ 
48	&	1-17-1	&	23:23:48.49 & +58:48:43.2 &  92.2	&	2.68	&	 161 (14) & $ -903~( 6)$	&	1.37	&	Fe	&	240 \\ 
49	&	2-14-2	&	23:23:49.34 & +58:48:32.1 &  95.9	&	2.81	&	 227 (17) & $-1554~( 9)$	&	1.34	&	Fe	&	251 \\ 
50	&	1-18-1	&	23:23:45.44 & +58:48:32.5 &  97.0	&	2.30	&	 273 ( 3) & $ +755~( 2)$	&	1.44	&	S	&	244 \\ 
51	&	2-14-1	&	23:23:49.15 & +58:48:28.4 &  97.2	&	2.79	&	 257 ( 6) & $ -207~( 2)$	&	1.32	&	Fe	&	250 \\ 
52	&	2-14-3	&	23:23:49.14 & +58:48:28.2 &  97.2	&	2.79	&	 246 ( 2) & $ +765~( 1)$	&	1.32	&	He	&	$\cdots$ \\ 
53	&	2-15-1	&	23:23:48.03 & +58:48:17.6 & 101.4	&	2.68	&	 303 (10) & $ -644~( 4)$	&	1.36	&	Fe	&	259 \\ 
54	&	2-15-2	&	23:23:47.96 & +58:48:16.1 & 102.0	&	2.67	&	 257 ( 2) & $+1187~( 1)$	&	1.36	&	He	&	$\cdots$ \\ 
55	&	1-19-1	&	23:23:43.98 & +58:48:19.2 & 103.4	&	2.16	&	 349 ( 9) & $ +753~( 5)$	&	1.41	&	S	&	270 \\ 
56	&	2-16-1	&	23:23:46.34 & +58:48:04.1 & 107.4	&	2.52	&	 230 (11) & $-1413~( 6)$	&	1.41	&	Fe	&	272 \\ 
57	&	2-17-2	&	23:23:41.93 & +58:48:08.8 & 110.3	&	1.95	&	 280 (15) & $-2052~( 7)$	&	1.43	&	S	&	$\cdots$ \\ 
58	&	1-20-1	&	23:23:42.88 & +58:48:05.4 & 110.5	&	2.09	&	 290 ( 7) & $-1651~( 4)$	&	1.39	&	S	&	277 \\ 
59	&	2-17-1	&	23:23:41.76 & +58:48:05.4 & 112.0	&	1.95	&	 320 (35) & $ -531~(14)$	&	1.42	&	S\tablenotemark{b}	&	275 \\ 
60	&	2-17-3	&	23:23:41.70 & +58:48:04.3 & 112.6	&	1.95	&	 405 (37) & $ +560~(18)$	&	1.46	&	S	&	$\cdots$ \\ 
61	&	3-23-1	&	23:23:37.55 & +58:48:13.4 & 115.3	&	1.40	&	 375 ( 5) & $ -674~( 3)$	&	1.68	&	S	&	282 \\ 
62	&	3-23-2	&	23:23:37.33 & +58:48:12.4 & 116.5	&	1.38	&	 300 ( 8) & $-4004~( 5)$	&	1.69	&	S	&	281 \\ 
63	&	2-18-1	&	23:23:44.54 & +58:47:31.3 & 120.9	&	2.53	&	 280 (22) & $-2023~(10)$	&	1.48	&	Fe	&	290 \\ 
64	&	1-22-1	&	23:23:43.68 & +58:47:33.3 & 121.6	&	2.42	&	 274 (40) & $-1725~(21)$	&	1.47	&	Fe	&	288 \\ 
65	&	2-19-1	&	23:23:40.78 & +58:47:28.5 & 128.7	&	2.16	&	 339 (13) & $  -44~( 7)$	&	1.42	&	S	&	293 \\ 
66	&	1-24-1	&	23:23:38.73 & +58:47:29.4 & 133.2	&	1.95	&	 436 (24) & $  +46~(13)$	&	1.43	&	S	&	296 \\ 
67	&	2-23-1	&	23:23:42.72 & +58:46:19.8 & 142.1	&	3.16	&	 197 ( 3) & $+1957~( 2)$	&	1.44	&	Fe	&	299 \\ 
\enddata
\tablecomments{
(1) knot serial number in this work; (2) Mask-Slit-number, where the number is 
the serial number of a knot detected in the slit; (3)--(4) coordinate of the emission feature in the slit;
(5) position angle measured counterclockwise from north to east
at the center of the SN explosion;
(6) radial distance from the center of the SN explosion;
(7)--(8) FWHM and central LSR velocity of the knot obtained from a weighted average of the emission lines.
The numbers in the parentheses are the $1\sigma$ statistical uncertainties
from single Gaussian fittings (see \S~\ref{sec-obs});
(9) hydrogen column density
derived from an {\it Herschel} SPIRE 250~\micron\ image;
(10) knot type:  S = S-rich knot, Fe = Fe rich knot, He=\hei\ knot. 
See \S~\ref{sec-res-cla} for an explanation about the classification of the knots.;
(11) Knot serial number in \citet{koo18}.  
}
\tablenotetext{a}{When there are multiple velocity components, 
they are labeled by ``A'' and ``B''. }

\tablenotetext{b}{\sii\ lines are detected, but \feii\ lines are not detected.}

\end{deluxetable}

%\clearpage

\begin{deluxetable*}{rllr}
\tabletypesize{\tiny}
\tablewidth{0pt}
\tablecolumns{4}
\tablecaption{Observed NIR Line Parameters of 67 Ejecta Knots \label{tab-lines}}
\tablehead{
\colhead{Knot}	&   \colhead{Line ID}	&
\colhead{$\lambda_{\rm rest}$}	&   \colhead{Observed Flux} \\
\colhead{No.}	&   \colhead{Transition ($l$--$u$)}	&
\colhead{($\micron$)}	&   \colhead{($10^{-17}$ erg s$^{-1}$ cm$^{-2}$)}   
}
\decimalcolnumbers
\startdata
 1	&	[S III] ~ $^{3}P_{2}$~--~$^{1}D_{2}$           	&	0.95337	&	   21.9 ( 2.8)	\\	%[2- 7-2]
 1	&	[C I] ~~~ $^{3}P_{1}$~--~$^{1}D_{2}$           	&	0.98268	&	$\cdot$ ( 2.4)	\\	%[2- 7-2]
 1	&	[C I] ~~~ $^{3}P_{2}$~--~$^{1}D_{2}$           	&	0.98530	&	   11.5 ( 1.7)	\\	%[2- 7-2]
 1	&	[S II] ~~ $^{2}D_{3/2}$~--~$^{2}P_{3/2}$       	&	1.02896	&	   15.0 ( 0.8)	\\	%[2- 7-2]
 1	&	[S II] ~~ $^{2}D_{5/2}$~--~$^{2}P_{3/2}$       	&	1.03233	&	   18.2 ( 0.0)	\\	%[2- 7-2]
 1	&	[S II] ~~ $^{2}D_{3/2}$~--~$^{2}P_{1/2}$       	&	1.03392	&	   11.3 ( 1.3)	\\	%[2- 7-2]
 1	&	[S II] ~~ $^{2}D_{5/2}$~--~$^{2}P_{1/2}$       	&	1.03733	&	    4.9 ( 0.0)	\\	%[2- 7-2]
 1	&	[S I] ~~~ $^{3}P_{2}$~--~$^{1}D_{2}$           	&	1.08241	&	   14.8 ( 2.0)	\\	%[2- 7-2]
 1	&	He I ~~~~ $^{3}S_{1}$~--~$^{3}P_{0,1,2}$       	&	1.08332	&	  257.0 ( 5.0)	\\	%[2- 7-2]
 1	&	[S I] ~~~ $^{3}P_{1}$~--~$^{1}D_{2}$           	&	1.13089	&	$\cdot$ ( 0.7)	\\	%[2- 7-2]
 1	&	[P II] ~~ $^{3}P_{2}$~--~$^{1}D_{2}$           	&	1.18861	&	$\cdot$ ( 1.0)	\\	%[2- 7-2]
 1	&	[Fe II] ~ $a^{6}D_{7/2}$~--~$a^{4}D_{5/2}$     	&	1.24888	&	$\cdot$ (21.3)	\\	%[2- 7-2]
 1	&	[Fe II] ~ $a^{6}D_{3/2}$~--~$a^{4}D_{1/2}$     	&	1.25248	&	$\cdot$ ( 2.4)	\\	%[2- 7-2]
 1	&	[Fe II] ~ $a^{6}D_{9/2}$~--~$a^{4}D_{7/2}$     	&	1.25702	&	    5.9 ( 1.8)	\\	%[2- 7-2]
 1	&	[Fe II] ~ $a^{6}D_{1/2}$~--~$a^{4}D_{1/2}$     	&	1.27069	&	$\cdot$ ( 3.3)	\\	%[2- 7-2]
 1	&	[Fe II] ~ $a^{6}D_{3/2}$~--~$a^{4}D_{3/2}$     	&	1.27913	&	$\cdot$ ( 2.1)	\\	%[2- 7-2]
 1	&	[Fe II] ~ $a^{6}D_{5/2}$~--~$a^{4}D_{5/2}$     	&	1.29462	&	$\cdot$ ( 2.4)	\\	%[2- 7-2]
 1	&	[Fe II] ~ $a^{6}D_{1/2}$~--~$a^{4}D_{3/2}$     	&	1.29813	&	$\cdot$ ( 4.5)	\\	%[2- 7-2]
 1	&	[Fe II] ~ $a^{6}D_{7/2}$~--~$a^{4}D_{7/2}$     	&	1.32092	&	$\cdot$ ( 1.2)	\\	%[2- 7-2]
 1	&	[Fe II] ~ $a^{6}D_{3/2}$~--~$a^{4}D_{5/2}$     	&	1.32814	&	$\cdot$ ( 0.9)	\\	%[2- 7-2]
 [5pt]
 1	&	[Fe II] ~ $a^{4}F_{9/2}$~--~$a^{4}D_{5/2}$     	&	1.53389	&	$\cdot$($\cdot$)	\\	%[2- 7-2]
 1	&	[Fe II] ~ $a^{4}F_{7/2}$~--~$a^{4}D_{3/2}$     	&	1.59991	&	$\cdot$($\cdot$)	\\	%[2- 7-2]
 1	&	[Si I] ~~ $^{3}P_{1}$~--~$^{1}D_{2}$           	&	1.60727	&	$\cdot$($\cdot$)	\\	%[2- 7-2]
 1	&	[Fe II] ~ $a^{4}F_{9/2}$~--~$a^{4}D_{7/2}$     	&	1.64400	&	$\cdot$($\cdot$)	\\	%[2- 7-2]
 1	&	[Si I] ~~ $^{3}P_{2}$~--~$^{1}D_{2}$           	&	1.64590	&	$\cdot$($\cdot$)	\\	%[2- 7-2]
 1	&	[Fe II] ~ $a^{4}F_{5/2}$~--~$a^{4}D_{1/2}$     	&	1.66422	&	$\cdot$($\cdot$)	\\	%[2- 7-2]
 1	&	[Fe II] ~ $a^{4}F_{7/2}$~--~$a^{4}D_{5/2}$     	&	1.67733	&	$\cdot$($\cdot$)	\\	%[2- 7-2]
 1	&	[Fe II] ~ $a^{4}F_{5/2}$~--~$a^{4}D_{3/2}$     	&	1.71159	&	$\cdot$($\cdot$)	\\	%[2- 7-2]
 1	&	[Fe II] ~ $a^{4}F_{3/2}$~--~$a^{4}D_{1/2}$     	&	1.74541	&	$\cdot$($\cdot$)	\\	%[2- 7-2]
\enddata
\tablecomments{
(1) knot serial number in this work; (2) line ID and its transition; 
(3) rest wavelength in vacuum; (4) observed line flux.
The numbers in the parentheses are the $1\sigma$ statistical uncertainties
from single Gaussian fittings (see S~\ref{sec-obs}).
}
\tablecomments{
This table is available in its entirety in a machine-readable form in the online journal.
}
\end{deluxetable*}

%\clearpage
\begin{deluxetable*}{llcc}
\tabletypesize{\scriptsize}
\tablewidth{0pt}
\tablecolumns{4}
\tablecaption{Spectral Classification of the 67 NIR Ejecta Knots and their Statistics\label{tab-group}}
\tablehead{
\colhead{Knot Group}	&   
\colhead{Spectral Properties}	& \multicolumn{2}{c}{Number of Knots} \\
\cline{3-4}
 & & NE jet area & Fe K plume area
\\
}
\decimalcolnumbers
\startdata
\textbf{S-rich Knots} &   
    strong \sii, \feii\ lines  &     31   &  13 \\
     &       strong \pii\ line (Fe K plume area); weak/no \pii\ line (NE jet area)   &  &  \\
     &        \siioverfeii\ $> 3.7$   &  &  \\
\hline
\textbf{Fe-rich Knots}   &   
    strong \feii\ line; weak or no \sii , \pii\ lines  &   1 &  15 \\ 
   &   \siioverfeii\ $< 3.7$    &  &  \\
\hline
\textbf{\hei\ Knots}    &  
    strong \hei\ line; weak or no \ci\ line   &  3 & 4 \\
    &  \heioverfeii\ $> 8$   &   &  \\
\hline
     \multicolumn{1}{c}{Total} & & 35 & 32  \\
\hline
\enddata
\tablecomments{(1) knot group; (2) line ratios; (3)--(4) number of knots in the NE jet and Fe K plume areas}
\end{deluxetable*}

%\clearpage
\begin{deluxetable*}{ccccrccrcccc}
\tabletypesize{\scriptsize}
\setlength{\tabcolsep}{3pt}
\tablecaption{Properties of Fe-rich Knots \label{tab-feknots}}
\tablewidth{0pt}
\tablecolumns{11} \tablehead{
\colhead{Knot} & \colhead{K2019 No} & \colhead{$\alpha$(J2000)} & \colhead{$\delta$(J2000)} &
\colhead{Size} & \colhead{$\mu$} &
\colhead{ $F_{{\rm [Fe\, II]} 1.644\, \mu{\rm m}}$ } &
\colhead{$v_r$} & \colhead{${F({\rm[S\, II]}) \over F({\rm [Fe\, II]})}$} &
\colhead{${F({\rm[P\, II]}) \over F({\rm [Fe\, II]})}$} &
\colhead{${F({\rm He\, I}) \over F({\rm [Fe\, II]})}$} & \colhead{Optical}\\
\colhead{No} & \colhead{ } &   &   &
\colhead{(arcsec)} & \colhead{(km s$^{-1}$)} &  
\colhead{(erg cm$^{-2}$ s$^{-1}$)} & 
\colhead{(km s$^{-1}$)} &&& & \colhead{Counterpart}
}
\decimalcolnumbers
\startdata
   23 &      197 & 23:23:57.399 &  58:50:30.36 &   2.26$\times$1.44 & $\cdots$ &    5.75E-16 &       -234 &        1.78 &   $\le$0.13 &   $\le$0.26 &        FMK \\
   39 &      235 & 23:23:49.208 &  58:49:04.65 &   1.85$\times$1.48 & $\cdots$ &    4.66E-16 &      -1286 &   $\le$1.72 &   $\le$0.10 &        0.71 &        FMK \\
   40 &      234 & 23:23:49.044 &  58:49:03.51 &   1.69$\times$1.44 & $\cdots$ &    5.17E-16 &      -1282 &   $\le$0.40 &   $\le$0.04 &        0.59 &        FMK \\
   42 &      232 & 23:23:48.305 &  58:48:59.55 &   1.85$\times$1.31 &     7291 &    4.81E-16 &      -1616 &   $\le$0.48 &   $\le$0.06 &        0.24 &        FMK \\
   43 & $\cdots$ & 23:23:48.336 &  58:48:57.60 &   0.77$\times$0.77 & $\cdots$ &    9.30E-17 &      -1987 &   $\le$1.13 &   $\le$0.20 &   $\le$0.60 &   $\cdots$ \\
   45 & $\cdots$ & 23:23:51.235 &  58:49:51.88 &   1.47$\times$1.39 & $\cdots$ &    3.82E-16 &      -1165 &   $\le$0.62 &   $\le$0.09 &        0.45 &   $\cdots$ \\
   46 &      241 & 23:23:48.617 &  58:48:54.88 &   3.29$\times$2.32 &     7221 &    2.98E-15 &      -1189 &   $\le$0.26 &   $\le$0.02 &        0.34 &        FMK \\
   47 &      239 & 23:23:48.106 &  58:48:53.21 &   3.41$\times$2.61 &     6989 &    2.89E-15 &      -1159 &   $\le$0.24 &   $\le$0.02 &        0.38 &        FMK \\
   48 &      240 & 23:23:48.253 &  58:48:43.33 &   1.26$\times$0.84 & $\cdots$ &    1.90E-16 &       -903 &   $\le$1.21 &   $\le$0.20 &   $\le$0.31 &        FMK \\
   49 &      251 & 23:23:49.109 &  58:48:32.44 &   1.21$\times$0.82 & $\cdots$ &    1.63E-16 &      -1554 &        3.13 &        0.35 &        0.86 &        FMK \\
   51 &      250 & 23:23:48.909 &  58:48:28.59 &   2.26$\times$1.76 & $\cdots$ &    9.13E-16 &       -207 &   $\le$0.39 &   $\le$0.13 &        8.02 &   $\cdots$ \\
   53 &      259 & 23:23:47.850 &  58:48:17.75 &   5.01$\times$2.89 &     7640 &    3.36E-15 &       -644 &        1.07 &        0.11 &        0.30 &        FMK \\
   56 &      272 & 23:23:46.136 &  58:48:04.65 &   1.46$\times$1.22 & $\cdots$ &    3.66E-16 &      -1413 &   $\le$0.85 &   $\le$0.09 &   $\le$0.16 &   $\cdots$ \\
   63 &      290 & 23:23:44.342 &  58:47:32.27 &   1.22$\times$0.89 & $\cdots$ &    1.87E-16 &      -2023 &   $\le$3.43 &   $\le$0.37 &   $\le$0.74 &   $\cdots$ \\
   64 &      288 & 23:23:43.483 &  58:47:33.93 &   5.86$\times$1.46 &     6580 &    1.10E-15 &      -1725 &   $\le$1.82 &   $\le$0.24 &   $\le$1.00 &        FMK \\
   67 &      299 & 23:23:42.554 &  58:46:21.52 &   2.54$\times$2.09 & $\cdots$ &    1.81E-15 &       1957 &   $\le$0.18 &   $\le$0.03 &        2.25 &        FMK \\
\enddata
\tablecomments{(1) knot serial number in this work; (2) knot serial numnber in \citet{koo18}; (3)--(4) central coordinate; (5) major and minor axis; (6) velocity of proper motion; 
(7) observed  [\ion{Fe}{2}] 1.644 $\mu$m flux; (8) radial velocity; (9)--(11) line ratios; (12) optical counterpart (see \S~\ref{sec-res-opt})}
\tablecomments{(3)--(7) Form \citet{koo18}}
\end{deluxetable*}

%\clearpage
\begin{deluxetable*}{lcccrrrrcc}
\tabletypesize{\scriptsize}
\setlength{\tabcolsep}{3pt}
\tablecaption{Ejecta Knots with High \heioverfeii\ \label{tab-heknots}}
\tablewidth{0pt}
\tablecolumns{10} \tablehead{
& \colhead{Knot} & \colhead{$v_r$} &
\colhead{ $F_{{\rm He I}\, 1.083\, \mu{\rm m}}$ } &
\colhead{${F({\rm[S\, II]}) \over F({\rm [Fe\, II]})}$} &
\colhead{${F({\rm He\, I}) \over F({\rm [Fe\, II]})}$} &
%\colhead{${F({\rm[P\, II]}) \over F({\rm [Fe\, II]})}$} &
%\colhead{${F({\rm[P\, II]}) \over F({\rm [Fe\, II]})}$} &
\colhead{${F({\rm [S\, II]}) \over F({\rm He\, I})}$} &
%\colhead{${F({\rm [C\, I]}\, 0.985\, \mu{\rm m}) \over F({\rm He\, I}\, 1.083\, \mu{\rm m})}$} &
\colhead{${F({\rm [C\, I]}) \over F({\rm He\, I})}$} &
\colhead{Knot} & \colhead{Optical} \\
& \colhead{No} &   \colhead{(km s$^{-1}$)} &
\colhead{(erg cm$^{-2}$ s$^{-1}$)} & &&&& \colhead{Type} & \colhead{Counterpart} \\
& \colhead{(1)} &  \colhead{(2)} &  \colhead{(3)} & \colhead{(4)} & \colhead{(5)} & \colhead{(6)} & \colhead{(7)} & \colhead{(8)} & \colhead{(9)} 
}
%\decimalcolnumbers
\startdata
\underline{He I knots} &    26 &       3241 &    3.89E-15 &    $\cdots$ &  $\ge$ 8.21 &  $\le$ 0.55 &  $\le$ 0.25 &         He &        FMK \\
  &    33 &      -3152 &    6.43E-15 &    $\cdots$ & $\ge$ 27.40 &  $\le$ 0.55 &        0.31 &         He &        FMK \\
  &    35 &       2314 &    2.22E-15 &    $\cdots$ & $\ge$ 14.50 &  $\le$ 0.39 &  $\le$ 0.57 &         He &   $\cdots$ \\
  &    38 &      -3312 &    1.34E-15 &    $\cdots$ & $\ge$ 14.70 &  $\le$ 0.47 &  $\le$ 0.23 &         He &        FMK \\
  &    41 &       -268 &    1.28E-15 &    $\cdots$ & $\ge$ 22.09 &  $\le$ 0.35 &  $\le$ 0.24 &         He &   $\cdots$ \\
  &    52 &        765 &    8.87E-15 &    $\cdots$ & $\ge$ 83.32 &  $\le$ 0.08 &        0.11 &         He &   $\cdots$ \\
  &    54 &       1186 &    1.05E-14 &    $\cdots$ & $\ge$ 77.55 &  $\le$ 0.07 &        0.07 &         He &   $\cdots$ \\
\hline
\underline{Other types of knots} &     1 &       2336 &    3.09E-14 &       17.21 &       73.23 &        0.24 &        0.07 &       S &        FMK \\
  &     5 &        123 &    1.80E-13 &       71.82 &       33.10 &        2.17 &        0.08 &          S &        FMK \\
  &    32 &      -3169 &    7.67E-15 & $\ge$ 46.20 & $\ge$ 54.43 &        0.85 &  $\le$ 0.09 &      S &   $\cdots$ \\
  &    51 &       -206 &    1.49E-14 &  $\le$ 0.39 &        8.02 &  $\le$ 0.05 &        0.13 &         Fe &   $\cdots$ \\
\enddata
\tablecomments{(1) knot serial number in this work; (2) radial velocity; 
(3) \ion{He}{1} 1.083 $\mu$m flux; (4)--(7) line ratios where [C I] = [C I] 0.985 \micron; (8) knot type; (9) optical counterpart (see \S~\ref{sec-res-opt})}
\end{deluxetable*}

%\clearpage
\begin{deluxetable*}{cccrrrrrr}
\tabletypesize{\scriptsize}
\tablewidth{0pt}
\tablecolumns{9}
\tablecaption{Physical Parameters of Three H Emission Features \label{tab-hi}}
\tablehead{
\colhead{Feature}	&   \colhead{$\alpha$(J2000)} & \colhead{$\delta$(J2000)}&
\colhead{P.A.}	&   \colhead{$d_{\rm rad}$}	&
\colhead{FWHM}	&   \colhead{$v_{\rm LSR}$} &
\colhead{F(Pa$\gamma$)} &   \colhead{F(Pa$\beta$)}  \\
\colhead{Name}	&   	&    &
\colhead{($\degr$)}	&   \colhead{($\arcmin$)}  &
\colhead{(\AA)}	&   \colhead{(\kms)}    &
\multicolumn{2}{c}{($10^{-17}$ erg s$^{-1}$ cm$^{-2}$)}  
}
\decimalcolnumbers
\startdata
 H1	&	23:23:45.81 & +58:48:52.5	&	 88.7	&	2.34	&	 8.8 (1.8)	&	$  -29~(17)$	&	$\cdots$ (1.1)	&	    5.1 (1.3)	\\	%[3-18-2]
 H2	&	23:23:49.14 & +58:48:28.0	&	 97.3	&	2.79	&	 8.7 (0.6)	&	$  -19~( 6)$	&	    6.1 (1.2)	&	   14.6 (1.4)	\\	%[2-14-4]
 H3	&	23:23:47.94 & +58:48:15.7	&	102.1	&	2.67	&	 7.0 (1.0)	&	$  -28~(10)$	&	$\cdots$ (0.8)	&	    6.6 (1.2)	\\	%[2-15-3]
\enddata
\tablecomments{
(1) feature name; (2)--(3) central coordinate of the feature;
(4)--(5); position angle and radial distance of the feature from the center of the SN explosion;
(6)--(7): FWHM and LSR velocity of the Pa$\beta$ line.
The numbers in the parentheses are their $1\sigma$ statistical uncertainties
from single Gaussian fittings.
They do not include the systematic uncertainties from
the absolute wavelength calibration ($\sim 12~\kms$; see \S~\ref{sec-obs}).;
(8)--(9) observed fluxes of Pa$\gamma$ and Pa$\beta$ lines.
The numbers in the parentheses are their $1\sigma$ statistical uncertainties
from single Gaussian fittings. They do not include the systematic uncertainties from
the absolute photometric calibration ($\sim 20$\%; see \S~\ref{sec-obs}).
}
\end{deluxetable*}

\end{document}